\documentclass[%
 reprint,
 amsmath,amssymb,
 aps,
 prl]{revtex4-2}
\usepackage{CJK}

\usepackage{mathtools,extarrows}
\usepackage{float,subfigure,tikz,tikzscale,graphbox}
\usepackage{shuffle}
\begin{document}
\begin{CJK*}{UTF8}{}
\CJKfamily{gbsn}
\title{Form factors from string amplitudes}
\author{Qu Cao$^{1,2,3}$}
\email{qucao@zju.edu.cn}
\affiliation{
$^{1}$Zhejiang Institute of Modern Physics, School of Physics, Zhejiang University, Hangzhou, Zhejiang 310027, China \\
$^{2}$Institute of Theoretical Physics, Chinese Academy of Sciences, Beijing 100190, China\\
$^{3}$ Joint Center for Quanta-to-Cosmos Physics, Zhejiang University, Hangzhou, Zhejiang 310058, China
}
\date{\today}

\begin{abstract}
In this letter, we propose a stringy model for $n$-point tree-level form factor with the off-shell operator in the scalar and gluon theories, from the bosonic string disk amplitude: $n$ open string states and $1$ closed string state scatter on the disk. In the field-theory limit ($\alpha'\to0$), the {\it stringy form factor} reduces to the form factor, helps us to investigate the hidden properties of the field-theory form factors, manifest the factorization and soft behaviors, and uncover more non-trivial relations between form factors and scattering amplitudes.


\end{abstract}

\maketitle
\end{CJK*}

\section{Introduction}

String-inspired methods have proven to be powerful tools for understanding field-theory observables, particularly in the study of scattering amplitudes~\cite{Kawai:1985xq,Bern:1990cu,Bern:1990ux,Bern:1991aq,Bern:1998sc,Bern:1998sv,Witten:2003nn,Cachazo:2013gna,Cachazo:2013iea,Cachazo:2013hca,Mason:2013sva,He:2018pol,He:2019drm}, also see the reviews~\cite{Bern:1992ad,Elvang:2013cua,Henn:2014yza,Travaglini:2022uwo,Mafra:2022wml,Berkovits:2022ivl}. In the field-theory limit, {\it i.e.}, when the inverse string tension $\alpha' \to 0$, open and closed string amplitudes reduce to field-theory scattering amplitudes for color-ordered and colorless theories, respectively~\cite{Green:1987sp}.  On one hand, string amplitudes inspire novel expressions and reveal hidden structures within scattering amplitudes~\cite{Kawai:1985xq,Witten:2003nn,Stieberger:2013wea,Stieberger:2014hba,Arkani-Hamed:2019mrd}. On the other hand, from the worldsheet perspective, string amplitudes exhibit intricate relations, such as Monodromy relations~\cite{Bjerrum-Bohr:2009ulz,Stieberger:2009hq}, which in turn imply nontrivial identities in field theory, including the Kleiss-Kuijf (KK)~\cite{Kleiss:1988ne} and Bern-Carrasco-Johansson (BCJ) relations~\cite{Bern:2008qj}.

As the cousin of scattering amplitudes, form factors have also attracted significant attention due to their theoretical importance~\cite{vanNeerven:1985ja,Maldacena:2010kp,Brandhuber:2010ad,Bork:2010wf,Brandhuber:2011tv}. The form factor is a matrix element between a local operator and on-shell asymptotic states, defined in momentum space through a Fourier transform:
\begin{equation}
    F_n^{\mathcal{O}}(1,2,\ldots,n;q) \equiv \int \mathrm{d}^D x\, e^{i q \cdot x} \langle 0| \mathcal{O}(x) |1,2,\ldots,n \rangle\,.
\end{equation}

Normally, the perturbative study of form factors relies on Feynman diagrams, which often obscure certain important properties, such as gauge invariance. The limitations of diagrammatic expansions have motivated the development of alternative tools that better manifest these hidden properties. Recently, the on-shell methods have great progress on form factors~\cite{Brandhuber:2010ad,Bork:2010wf,Brandhuber:2011tv,Brandhuber:2012vm,Brandhuber:2014ica,Boels:2012ew,Bork:2014eqa,Frassek:2015rka,Bork:2016hst,Bork:2016xfn,Yang:2016ear,Bork:2017qyh,Bianchi:2018peu}, also see the reviews~\cite{Nandan:2018hqz,Yang:2019vag,Lin:2022jrp,Lin:2023rwe}. However, string-inspired methods for form factors remain largely unexplored, particularly in the absence of supersymmetry. The string-inspired method serves as a UV-completion model for field-theory form factors. More importantly, string-inspired formulas often reduce the expression to a single, unified integral that reveals these underlying symmetries and properties.

In this letter, we take the first step toward developing a {\it stringy form factor}, a stringy model for tree-level form factors, motivated by the structure of open $\&$ closed string disk amplitudes~\footnote{To be precise, while we believe that a stringy model of form factors may exist on higher-genus Riemann surfaces, we restrict our focus here to the simplest case: the disk topology, which corresponds to the tree-level.}. The stringy form factor manifests the pole structures/factorizations and soft behaviors, and the new hidden properties: 2-split. More intersecting, there is a new non-trivial relation between form factors and scattering amplitudes in the field-theory can be derived from the aspect of the stringy form factors.

\section{ a stringy model on the disk}

A main difference between the form factor and scattering amplitude is the off-shell leg in the form factor. The off-shell leg can be viewed as a massive scalar leg, which turns the $n$-point form factor as a $(n+1)$ scattering amplitude with a colorless massive scalar. In the context of string amplitude, it is hard to consider the off-shell vertex operators~\footnote{Some literature studies the one massive string amplitudes~\cite{Guillen:2021mwp,Kashyap:2023cdi,Kashyap:2024qor,Mafra:2024fiy}.}, but in the context of the open $\&$ closed string disk amplitudes(considered as the interaction between the boundary and bulk fields from the unstable D-brane~\cite{Klebanov:1995ni,Gubser:1996wt,Garousi:1996ad,Hashimoto:1996bf,Hashimoto:1996kf,Sen:2003bc,Sen:2003xs}), the off-shell momentum can exist because the momentum conservation only exists in the longitudinal direction of the D-brane.

Inspired by open $\&$ closed string scattering on the disk~\cite{Stieberger:2009hq,Stieberger:2015vya}\footnote{If we actually choose the metric matrix $D^{\mu\nu}$ (the same notation as in~\cite{Stieberger:2015vya}) to be a specific value, we cannot realize the kinematic condition for the form factor ($q^2 \neq 0$).  Here we actually generalize the kinematics part of disk string amplitude to the stringy form factor in~\eqref{eq:stringy ff def} by mapping the kinematics
$2q^{1} \cdot q^{2} \to q^2,
2p_i \cdot q^{1} = 2p_i \cdot q^{2} \to -p_i \cdot q$, ($q^{1}, q^{2}$ are defined in~\cite{Stieberger:2015vya}).

This map cannot be realized by any choice of the matrix $D$.}, we propose a stringy model for the UV completion of the tree-level form factor for scalar and gluon theories. The stringy form factor $\mathcal{F}^{\mathcal{O}}_n(1,2,\dots,n;q)$ is

\begin{equation}\label{eq:stringy ff def}
\mathcal{F}^{\mathcal{O}}_n(1,2,\dots,n;q)=\int {\rm d}\mu_{n;0}\times \mathcal{I}_{n}^{\mathcal{O}} \times\text{KN}(1,\ldots,n;q)
\end{equation}

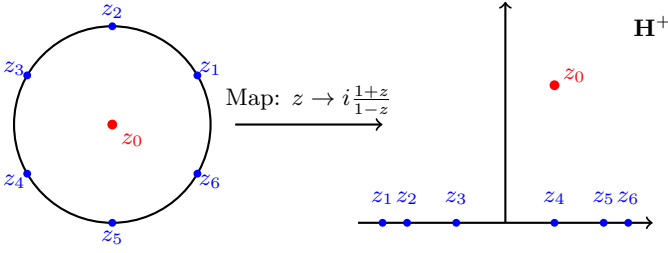
\begin{figure}
    \centering
    \begin{tikzpicture} [scale=0.65]
    \draw[thick] (-4,0) circle (2);
    
    \foreach \angle [count=\i] in {30, 90, 150, 210, 270, 330} {
        \fill[blue] ({-4+2*cos(\angle)}, {2*sin(\angle)}) circle (0.08);
        \node[blue, font=\small] at ({-4+2.3*cos(\angle)}, {2.3*sin(\angle)}) {$z_{\i}$};
    }

    \fill[red] (-4,0) circle (0.1);
    \node[red, below right] at (-4,0) {$z_{0}$};

    \draw[->,thick] (1,-2) -- (7,-2); 
    \draw[->,thick] (4,-2) -- (4,2.5); 

    \foreach \x [count=\i] in {1.5, 2, 3, 5, 6,6.5} {
        \fill[blue] (\x, -2) circle (0.08);
        \node[blue, above] at (\x, -1.8) {$z_{\i}$};
    }

    \fill[red] (5, 0.8) circle (0.1);
    \node[red, right] at (5,1) {$z_{0}$};
\node[black] at (7,2) {$\bold{H}^{+}$};
    \draw[->, thick] (-1.5,0) -- (1.5,0);
    \node[above] at (0,0) {Map: $z \rightarrow i \frac{1+z}{1-z}$};

\end{tikzpicture}

    \caption{Open(blue) and closed(red) string vertex positions on the disk.  The unit disk $\mathcal{D}=\{z \in \mathbf{C}||z| \leq 1\}$ on the left, can be conformally mapped to the upper half plane $\mathbf{H}_{+}=\{z \in \mathbf{C} \mid \operatorname{Im}(z) \geq 0\}$ on the right by the map: $z \rightarrow i \frac{1+z}{1-z}$.}
    \label{fig:disk}
\end{figure}

This stringy integral represents a UV completion of the $n$-point form factor involving an off-shell momentum $q$ and an operator $\mathcal{O}$, which satisfies the momentum conservation condition $q=\sum_{i=1}^{n} p_{i}$. The disk hosts $n$-point open-string vertex operators inserted in a specific ordering $\Gamma$(for simplicity, we adopt the canonical ordering $z_{1}< z_{2}<\ldots < z_{n}$ here). Additionally, a single closed-string vertex operator is inserted in the interior of the disk, which is mapped to the upper half-plane $\bold{H}^{+}=\text{Im}(z)\geq 0$, as illustrated in Figure~\ref{fig:disk}.

The measure $\int {\rm d}\mu_{n;0}$ is defined as,

\begin{equation}
  \int {\rm d}\mu_{n;0}\equiv\int_{\Gamma}\frac{\mathrm{d}z_1\dots\mathrm{d}z_n}{\mathrm{vol}\text{ SL}(2,\mathbb{R})} \int_{\bold{H}^{+}}\frac{\mathrm{d}z_{0}\mathrm{d}{\bar z}_{0}}{(z_{0,{\bar 0}})^{2}}\,,
\end{equation}
with the usual $\text{ SL}(2,\mathbb{R})$ symmetry on the disk. The universal part is the Koba-Nielsen factor~\cite{Koba:1969rw}, 
with the usual $\text{ SL}(2,\mathbb{R})$ symmetry on the disk. The universal part is the Koba-Nielsen factor~\cite{Koba:1969rw}, 
\begin{equation}
  \text{KN}(1,\ldots,n;q)=\prod_{i<j}z_{i,j}^{2\alpha'p_i\cdot p_j} \prod_{i}(z_{i,0}z_{i,{\bar 0}})^{-\alpha'p_i\cdot q}(z_{0,{\bar 0}})^{\alpha'q^2}\,,
\end{equation}
where $z_{a,b}=z_{a}-z_{b},z_{a,{\bar 0}}=z_{a}-{\bar z}_{0}$, and one can check this factor contributes zero $\text{ SL}(2,\mathbb{R})$ weights for each $z_{i}$ and $z_{0}/{\bar z_{0}}$. The integrand $\mathcal{I}_{n}^{\mathcal{O}}$ encodes the information of the operator $\mathcal{O}$ and the $n$-point external asymptotic states. Here we already pull out $(z_{0}-{\bar z_{0}})^{-2}$ from the integrand $\mathcal{I}_{n}^{\mathcal{O}}$, so the $\text{ SL}(2,\mathbb{R})$ weights  of $\mathcal{I}_{n}^{\mathcal{O}}$ are $-2$ for each $z_{i}$, and $0$ for $z_{0}/{\bar z_{0}}$. The $\mathcal{I}_{n}^{\mathcal{O}}$ still can depend on $z_{0}/{\bar z_{0}}$. We list some integrands, as follows.

The simplest integrand is the Parke-Taylor factor~\cite{Parke:1986gb}, which from the Wick contraction of the Kac-Moody current $\mathcal{J}_{a}(z_{i})$~\cite{Broedel:2013tta,Carrasco:2016ldy,Carrasco:2016ygv}. This integrand corresponds to the operator $\mathcal{O}=\frac{1}{2}\text{Tr}((\partial \phi)^{2})+\text{Tr}(\phi^{3})$, the Lagrangian $\mathcal{L}_{\phi^{3}}$ for the $\text{Tr}(\phi^{3})$ theory. The $n$-point external states are scalars.
\begin{equation}
    \begin{aligned}
       \mathcal{I}_{n}^{\mathcal{L}_{\phi^{3}}}=\frac{1}{z_{1,2}z_{2,3}\dots z_{n,1}}=\text{PT}(1,2,\ldots,n)\,.
    \end{aligned}
\end{equation}

The next integrand from the gluon vertex operators, which contract to the familiar building blocks $V_{i}$ and $W_{i,j}$~\cite{Green:1987sp}. The $W_{i,j}$ is the same as usual in the context of the string amplitude, but now the $V_{i}$ is modified due to the contraction between the vertices $z_{i}$ and $z_{0}/{\bar z_{0}}$. The definitions of $V$ and $W$ are,
\begin{equation}
\begin{aligned}
  &V_{i}:= \sum_{j\neq i}^n \frac{2\epsilon_i \cdot p_j}{z_{i,j}}-\frac{\epsilon_i \cdot q}{z_{i,0}}-\frac{\epsilon_i \cdot q}{z_{i,{\bar 0}}}\\
  &=\sum_{j\neq i}^n \epsilon_i \cdot p_j\left(\frac{z_{j,0}}{z_{i,j}z_{i,0}}+\frac{z_{j,{\bar 0}}}{z_{i,j}z_{i,{\bar 0}}}\right), \quad W_{i,j}:=\frac{\epsilon_i \cdot \epsilon_j}{\alpha' z_{i,j}^2} 
\end{aligned}   
\end{equation}
where we have used the momentum conservation.

The operator $\mathcal{O}=\text{Tr}(F^{2})$, and the external states are gluons in the pure Yang-Mills theory.

\begin{equation}
    \begin{aligned}
\mathcal{I}_{n}^{\text{Tr}(F^{2})}=\sum_{r=0}^{\lfloor n/2 \rfloor{+}1} \sum_{\{g,h\}, \{l\}} \prod_{s}^r W_{g_s, h_s} \prod_{t}^{n-2r} V_{l_t}\,,
    \end{aligned}
\end{equation}
where we have a summation over all partitions of $\{1,2,\cdots, n\}$ into $r$ pairs $\{g_s, h_s\}$ and $n-2r$ singlets $l_t$, each summand given by the product of $W$'s and $V$'s, which are from the Wick contraction of the gluon vertex operators $\epsilon_{i}\cdot \partial X(z_{i})$. 

For $n=2$, the integrand is $\mathcal{I}_{2}^{\text{Tr}(F^{2})}=V_{1}V_{2}+W_{1,2}$ . For $n=3$, the integrand is $\mathcal{I}_{3}^{\text{Tr}(F^{2})}=V_{1}V_{2}V_{3}+W_{1,2}V_{3}+W_{2,3}V_{1}+W_{1,3}V_{2}$ .

The $\text{Tr}(F^{2})$ form factor can expand as the sum of $\text{Tr}(\phi^{2})$ form factors~\cite{Dong:2022bta}, similar universal expansion studied for amplitudes at tree and one-loop level~\cite{Dong:2021qai,Cao:2024olg}. With the help of this non-trivial relation, we can construct a stringy model for the $\text{Tr}(\phi^{2})$ form factors, taking the coefficient from $\mathcal{I}_{n}^{\text{Tr}(F^{2})}$.
\begin{equation}
       \mathcal{I}_{n}^{\text{Tr}(\phi^{2})}=C_{1,2}C_{2,3}\ldots C_{n,1},\quad C_{i,j}=\frac{z_{j,0}}{z_{i,j}z_{i,0}}+\frac{z_{j,{\bar 0}}}{z_{i,j}z_{i,{\bar 0}}}\,,
\end{equation}
where we take the coefficient of $(\epsilon_1\cdot p_2\epsilon_2\cdot p_3\ldots \epsilon_n\cdot p_1)$  from $\mathcal{I}_{n}^{\text{Tr}(F^{2})}$ to obtain $\mathcal{I}_{n}^{\text{Tr}(\phi^{2})}$, due to the expansion relation~\cite{Dong:2022bta}. One can also construct other $\mathcal{I}_{n}^{\text{Tr}(\phi^{2})}$ form factors with more external gluon states from the expansion relation. In this letter, we focus on the first two stringy form factors.

In summary, we already choose the closed-string vertex operator inserted to be the Kac-Moody current $ \mathcal{J}_{a}(z_0) \mathcal{J}_{a}(\bar{z}_0)$ to form the $1/(z_0 - \bar{z}_0)^2$ in the integrand, and change the open-string vertex operators, such as $\mathcal{J}_{a}(z_i)$, and $\epsilon_{i}\cdot \partial X(z_{i})$ to define the integrand of stringy form factors~\eqref{eq:stringy ff def}. 

In the field-theory limit $\alpha'\to0$, the stringy form factors $\mathcal{F}^{\mathcal{O}}_n(1,2,\dots,n;q)$ reduces to form factor $F_n^{\mathcal{O}}(1, \ldots, n;q)$ in the field theory.
\begin{equation}\label{eq:field-theory}
    \lim_{\alpha'\to0}\mathcal{F}^{\mathcal{O}}_n(1,2,\dots,n;q)=F_n^{\mathcal{O}}(1,2,\dots,n;q )\,.
\end{equation}

We list some field-theory form factor results at the low multiplicity.

\begin{equation}
    \begin{aligned}
 &F_2^{\mathcal{L}_{\phi^{3}}}(1,2)=s_{1,2}\,,\quad F_2^{\text{Tr}(F^{2})}(1,2)=\text{tr}(1,2)\,,\\
 &F_3^{\mathcal{L}_{\phi^{3}}}(1,2,3)=\frac{(s_{1,2}+s_{2,3})(s_{1,2}+s_{1,3})(s_{1,3}+s_{2,3})}{s_{1,2}s_{1,3}s_{2,3}}\,,
    \end{aligned}
\end{equation}
where we have defined the shorthands $\text{tr}(\alpha_1,\ldots,\alpha_m)=\text{Tr}(f_{\alpha_1} \cdots f_{\alpha_m}):=(f_{\alpha_1})_{\mu_1}^{\; \mu_2}\cdots(f_{\alpha_m})_{\mu_m}^{\; \mu_1}$, and the field strength
$f_i^{\mu \nu}=k_i^\mu \epsilon_i^\nu -k_i^\nu \epsilon_i^\mu$.

In the following sections, to examine whether our stringy form factors have the correct field-theory limit~\eqref{eq:field-theory}, we analyze their pole structures and factorization properties, which should match those of field-theory form factors. Additionally, to clarify why the operators always correspond to the Lagrangian, we investigate the soft limit of the stringy form factors.
\section{Factorization}
We now briefly discuss the factorization of the stringy form factor. For simplicity, we focus on the stringy form factor associated with the massless poles, which survive in the field-theory limit. The factorization of the form factor at massless poles can be expressed as a product of a sub form factor and a scattering amplitude.
\begin{equation}
    \mathop{\rm Res}_{s_{1,\ldots,i}=0}F_n^{\mathcal{O}}(1,\ldots,n)=A_{i+1}(1,\ldots,i,I)F_{n{-}i+1}^{\mathcal{O}}({-}I,\ldots,n)
\end{equation}
with $s_{1,\ldots,n}=(p_{1}+\ldots+p_{i})^{2}=p_{I}^{2}\equiv s_{I}$, and we have omitted the state sum of the intermediate particle $I$.

All the massless poles are denoted by $s_{I}$, where $I$ is an ordered subset of $\{1,2,\ldots,n\}$, and the total number of such poles is $n(n-2)$. These poles arise from the pinching behavior of the corresponding punctures $z_i$ for $i \in I$, see e.g.,~\cite{Green:1987sp,Cachazo:2012pz,Cachazo:2013gna,Baadsgaard:2015voa}. For example, when $z_{1}$ pinches $z_{2}$, \textit{i.e.} when $z_{1} \to z_{2}$, the stringy integral will have a massless pole at $s_{1,2}$. More generally, we choose $z_{i}=\epsilon u_{i}$, with $i=1,2,\ldots,i$. With gauge fixing $u_{1}=0,u_{i}=1$, when $\epsilon\to0$, the integrand factorizes. For the PT factor, we have
\begin{equation}
    \text{PT}\to \epsilon^{1-i} \frac{1}{u_{1,2}z_{2,3}\dots u_{i,I}u_{I,1}} \times \frac{1}{z_{-I,i+1}z_{i+1,i+2}\dots z_{n,-I}}\,,
\end{equation}
with the added gauge fixing $u_{I}=\infty$, $z_{-I}=0$. Combined with the new measure and the KN factor, the integral of ${\rm d}\epsilon$ gives the pole of $s_{1,\ldots,i}$: $\int \epsilon^{\alpha' s_{I}-1}{\rm d}\epsilon = \frac{1}{\alpha' s_{I}}$, which represents the factorization of the stringy form factors. A similar derivation works for the integrand $\mathcal{I}_{n}^{\text{Tr}(F^{2})}$.

\section{Soft limit: $q\to0$}
The reason why all stringy form factors corresponding to the Lagrangian operator can be explained through the soft limit for the off-shell momentum. Let us now consider the soft limit: $q\to0$.

As far as we know, only when the local operator is the Lagrangian of the theory, the form factor reduces to the corresponding amplitude in the soft limit, as $q\to0$ (see an example in $\mathcal{N}=4$ SYM~\cite{Bork:2010wf}).

\begin{equation}
\begin{aligned}
    &F_n^{\mathcal{L}_{\phi^{3}}/\text{Tr}(F^{2})}(1, \ldots, n)\xrightarrow{q\to0}A_{n}^{\phi^3/\text{YM}}(1,\ldots, n)\,.
\end{aligned}
\end{equation}
This relation can be easily derived from the Lagrangian insertion.
\begin{equation}
    \int \mathrm{d}^D x\langle 0| \mathcal{L}(x)\left|1,\ldots,n \right\rangle=g\frac{\partial A_{n}}{\partial g}\propto A_{n}\,.
\end{equation}
where $g$ denotes the coupling constant in the theory. The soft limit on $q$, can generalize to the stringy level.
\begin{equation}\label{eq:soft}
\begin{aligned}
    &\mathcal{F}_n^{\mathcal{L}}(1, \ldots, n)\xrightarrow{q\to0}\mathcal{A}_{n}(1,\ldots, n)
\end{aligned}
\end{equation}
where $\mathcal{A}$ denotes the stringy integral for the scattering amplitudes.

One can consider the relation in this way. The limit is that we take $\alpha' \to 0$ first, then take $q \to 0$. The leading term in the field-theory limit has the correct soft behavior. But if we just want to argue the conjecture~\eqref{eq:soft}, we switch the sequence of these two limits. In the integrand level, if we ignore the integral on ${\rm d} z_0 {\rm d} z_{\bar{0}}$, which will be divergent when $q \to 0$ at first, one can introduce a regulator to make the integral finite.
Other remaining integrands reduce to the corresponding integrands in the string amplitude when $q \to 0$. The $\text{KN}(1,\ldots,n;q) \to \text{KN}(1,\ldots,n) = \prod_{i<j} z_{i,j}^{2\alpha' p_i \cdot p_j}$. Also, those $V$'s and $W$'s. Then, $\mathcal{F}_n^{\mathcal{L}_{\phi^{3}}}(1, \ldots, n)$ reduces to the stringy integrand with the Parke-Taylor integrand, which is the stringy amplitude for $\text{Tr}(\phi^3)$ theory~\cite{Broedel:2013tta,Arkani-Hamed:2019mrd}. Additionally, $\mathcal{F}_n^{\text{Tr}(F^2)}(1, \ldots, n)$ reduces to the bosonic string amplitude for YM theory~\cite{Green:1987sp}. These explanations clarify why all the stringy form factors considered here correspond to the Lagrangian operators.

\section{2-split}
From the above discussion, we have believed that the stringy form factors~\eqref{eq:stringy ff def} are meaningful in the sense of correct field-theory limits. Now we use the stringy form factors to explore the hidden properties of the form factors.

A non-trivial behavior called $2$-split has been studied in the string/particle amplitudes~\cite{Cao:2024gln,Cao:2024qpp}, which can be derived from the string integrand. The 2-split can derive the zeros and the factorization near zeros, which have been considered in the context of scattering amplitudes/string amplitudes~\cite{Arkani-Hamed:2023swr,Cao:2024gln,Cao:2024qpp,Rodina:2024yfc,Zhou:2024ddy}, and other interesting splits/factorization~\cite{Cachazo:2021wsz,Arkani-Hamed:2024fyd,Zhang:2024iun,Zhang:2024efe,GimenezUmbert:2025ech}. Now, with this new stringy model for form factor, one would ask if there is a $2$-split for these form factors. The answer is yes, but not all stringy form factors. 

{\it 2-split} kinematics for stringy form factor as follows: pick $3$ particles, $i,j,k$, and divide the remaining legs into two sets $A, B$, {\it i.e.} $A\cup B=\{1,\cdots, n\}/\{i,j,k\}$, then we demand\footnote{One of the subset $A$ or $B$ can be empty.},
\begin{align}
    s_{a,b} = 0, s_{a,q}=0,\qquad \forall a \in A, b \in B\,,
\end{align} 
where $s_{a,b}:=2 p_a \cdot p_b$, and  $s_{i,q}:=2 p_i \cdot q$. 

The simplest case for $2$-split is $\mathcal{F}^{\mathcal{L}_{\phi^{3}}}_n(1,2,\dots,n;q)$. Because the integrand $\mathcal{I}_{n}^{\mathcal{O}}$ is $\text{PT}(1,2,\ldots,n)$, which can naturally split as we proved in~\cite{Cao:2024gln,Cao:2024qpp}. The $\text{KN}(1,\ldots,n;q)$ for the stringy form factor also splits as a similar proof in ~\cite{Cao:2024gln,Cao:2024qpp}. We summarize the $2$-split, with $s_{a,b}=s_{a,q}=0$:
\begin{equation}
\begin{aligned}
   \mathcal{F}^{\mathcal{L}_{\phi^{3}}}_n(1,\dots,n;q) \to \mathcal{A}^{\phi^{3}}_{|A|+3}(i,A,j; \kappa)  \mathcal{F}^{\mathcal{L}_{\phi^{3}}}_{|B|+3} (j,B,i; \kappa',q)\\
\end{aligned}
\end{equation}
where the $\mathcal{A}^{\phi^{3}}$ denotes the open-string amplitude for massless $\phi^{3}$. The off-shell leg with momentum
$p_\kappa=-\sum_{a\in A} p_a -p_i-p_j$, and $p_{\kappa'}=q-\sum_{b\in B} p_b-p_i-p_j$.  

For $2$-split for $\mathcal{F}^{\text{Tr}(F^{2})}_n(1,2,\dots,n;q)$, we need more conditions as similar as the open-string amplitude. For $s_{a,b}=s_{a,q}=0$, we demand:
\begin{equation}\label{eq_pol}
\epsilon_a \cdot \epsilon_{b'}=0\,, \quad p_a \cdot \epsilon_{b'}=0\,,\quad \epsilon_a \cdot p_b=0\,,\quad  \epsilon_a \cdot q=0 ,
\end{equation}
for $a\in A, b \in B$ and $b' \in B \cup \{i,j,k\}$. Those $V$ and $W$'s split as similar as bosonic string amplitudes~\cite{Cao:2024gln,Cao:2024qpp},

\begin{equation}\label{bos_string}
\mathcal{F}^{\text{Tr}(F^{2})}_n \to {\cal J}^{\rm mixed}_{|A|+3} (i^\phi, A, j^\phi; \kappa^\phi) \mathcal{F}^{\text{Tr}(F^{2})}_{|B|+3} (j, B, i; \kappa',q)_\mu \epsilon_{k}^\mu\,,  
\end{equation}
where ${\cal J}_{|A|+3}^{\rm mixed}$ is a current with $|A|$ gluons (in $A$) and $3$ $\phi's$.

\begin{figure}
    \centering
    \begin{tikzpicture}[scale=0.6]
    \draw[thick] (-4,0) circle (2);

    \fill[red] (-4,0) circle (0.1);
    \node[red, below right] at (-4,0) {$z_{0}$};

    \foreach \angle [count=\i] in {45, 135, 225, 315} {
        \fill[blue] ({-4+2*cos(\angle)}, {2*sin(\angle)}) circle (0.08);
        \node[blue, font=\small] at ({-4+2.3*cos(\angle)}, {2.3*sin(\angle)}) {$z_{\i}$};
    }
    \draw[->, thick] (-1.5,0) -- (1.5,0);
    \node[above] at (0,0) {$i=3,j=1,k=2$};
    \node[below] at (0,0) {$s_{4,q}=0$};
    
    \draw[thick] (4,2) circle (1.5);

    \fill[red] (4,2) circle (0.1);
    \node[red, below right] at (4,2) {$z_{0}$};

 \fill[blue]({4+1.5*cos(45)},{2+1.5*sin(45)}) circle (0.1);
    \node[blue,right] at ({4+1.7*cos(45)},{2+1.7*sin(45)}) {$z_{1}$};
    \fill[blue]({4+1.5*cos(135)},{2+1.5*sin(135)}) circle (0.1);
    \node[blue,left] at ({4+1.7*cos(135)},{2+1.7*sin(135)}) {$z_{\kappa}$};
    \fill[blue]({4+1.5*cos(225)},{2+1.5*sin(225)}) circle (0.1);
    \node[blue,left] at ({4+1.7*cos(225)},{2+1.7*sin(225)}) {$z_{3}$};

    \draw[thick] (4,-2) circle (1.5);

   \fill[blue]({4+1.5*cos(45)},{-2+1.5*sin(45)}) circle (0.1);
    \node[blue,right] at ({4+1.7*cos(45)},{-2+1.7*sin(45)}) {$z_{1}$};
    \fill[blue]({4+1.5*cos(135)},{-2+1.5*sin(135)}) circle (0.1);
    \node[blue,left] at ({4+1.7*cos(135)},{-2+1.7*sin(135)}) {$z_{\kappa'}$};
    \fill[blue]({4+1.5*cos(225)},{-2+1.5*sin(225)}) circle (0.1);
    \node[blue,left] at ({4+1.7*cos(225)},{-2+1.7*sin(225)}) {$z_{3}$};
    \fill[blue]({4+1.5*cos(315)},{-2+1.5*sin(315)}) circle (0.1);
    \node[blue,right] at ({4+1.7*cos(315)},{-2+1.7*sin(315)}) {$z_{4}$};
    
\end{tikzpicture}

    \caption{The example for 2-split}
    \label{fig:2split}
\end{figure}
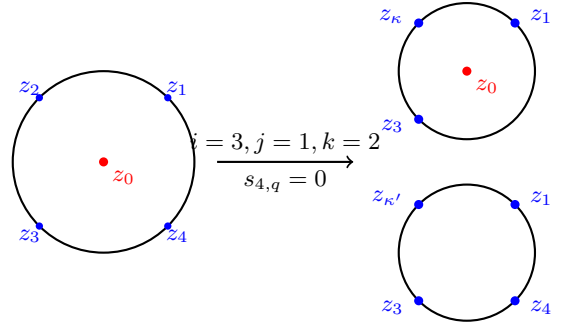

For example, we let $i=3,k=2,j=1$, and set the conditions:$s_{4,q}=0$($s_{2,4}=-s_{1,4}-s_{3,4}$) for form factor, depicted in the figure~\ref{fig:2split}.

$F_4^{\mathcal{L}_{\phi^{3}}}(1,2,3,4)\to$:
\begin{equation}
\begin{aligned}
   &  \frac{\left(s_{1,4}{+}s_{3,4}\right) \left(2p_{1}\cdot p_{2,3,4}\right) \left(s_{1,2}{-}s_{1,4}{+}s_{2,3}{-}s_{3,4}\right) \left(2p_{3}\cdot p_{1,2,4}\right)}{s_{1,4}s_{3,4} \left(s_{2,3}{-}s_{1,4}\right) \left(s_{1,2}{-}s_{3,4}\right)  \left(s_{1,3}{+}s_{1,4}{+}s_{3,4}\right)} \\
   &=A^{\phi^{3}}_{4}(3,4,1;\kappa) F^{\mathcal{L}_{\phi^{3}}}_{3} (1,3; \kappa',q)
\end{aligned}
\end{equation}
where $A^{\phi^{3}}$ is the field-theory massless $\phi^3$ amplitude, and the momentum $p_\kappa=-p_{4} -p_3-p_1$, and $p_{\kappa'}=q-p_1-p_3=p_{2}+p_{4}$, with the shorthand $p_{i,j,k}=p_{i}+p_{j}+p_{k}$.

We need to comment here that the stringy form factor $\mathcal{F}^{\text{Tr}(\phi^{2})}_n(1,2,\dots,n;q)$ can not split, because the integrand $\mathcal{I}_{n}^{\text{Tr}(\phi^{2})}=C_{1,2}C_{2,3}\ldots C_{n,1}$ depends on $z_{0}/{\bar z}_{0}$ and can not split under any $s_{i,j}/s_{i,q}=0$ constraint.

This is the first time we have found new behaviors in form factors. Furthermore, by imposing additional conditions similar to those in amplitudes, we can determine the zeros and the factorizations near zeros for form factors. We will not elaborate further on these aspects here. But, it would be interesting to unify form factors and scattering amplitudes in the stringy languages, as they share the same 2-split behavior. We leave this intriguing direction for future exploration.
\section{Relations between form factors and amplitudes}
In order to calculate the stringy form factor, the useful method is expanding the stringy form factor~\eqref{eq:stringy ff def} to a basis of open string amplitudes, based on the method~\cite{Stieberger:2009hq,Stieberger:2015vya}. We can analytically deform the image part of $z_{0}/z_{\bar 0}$ to the real line, with the gauge fixing for other puncture $z_{i}$'s. The integrand now becomes the $n{+}2$-point open string amplitude with different phase factors. After combining the Monodromy relations, the stringy form factor can be expanded as the sum of the $(n-1)!$ minimal basis open string amplitudes\footnote{See the ref~\cite{Stieberger:2009hq,Stieberger:2015vya} for the detailed derivation and discussion.} (Actually, only $2^{n-2}$ open string amplitudes contribute in the expansion). The expansion is
\begin{equation}
\begin{aligned}
&\mathcal{F}^{\mathcal{O}}_n(1,\ldots,n;q)=\sum_{j=2}^{n}(-1)^j \sin \left(\pi s_{j, n{+}1}\right) e^{\pi i(-1)^j s_{j, n{+}1}}\\
&(-1)^{n+2} e^{-\pi i\left(s_{1, n{+}1}+s_{2, n{+}2}\right)} \sum_{\rho } \mathcal{M}(\rho) \mathcal{A}(1, \rho,  n{+}1, n{+}2)
\end{aligned}
\end{equation}
where $\rho \in\left\{\alpha\shuffle\beta^T, j\right\}$, $\alpha=\{2, \ldots, j{-}1\}, \beta=\{j{+}1, \ldots, n\}$, $\beta^T$ is the reversal of ordered set $\beta$, and $\mathcal{M}(\rho)$ is the phase factor depending on the ordered set $\rho$, the expansion equals to the  equation~$(21)$ in~\cite{Stieberger:2015vya}, once set the kinematic conditions $s_{i,n+1}=s_{i,n+2}=-\frac{s_{i,q}}{2}=-p_{i}\cdot q$, $s_{n+1,n+2}=q^2$. Note that the string amplitudes $\mathcal{A}(1, \rho,  n{+}1, n{+}2)$ only change the integral contour but have the same integrand with stringy form factors by identifying $z_{0}=z_{n+1},{\bar z}_{0}=z_{n+2}$.

Here we only focus on the field-theory limit, when $\alpha'\to0$, the expansion relation becomes simpler:
\begin{equation}\label{eq:relation}
\begin{aligned}
F^{\mathcal{O}}_n(1,\ldots,n;q)\propto\sum_{j=2}^{ n}(-1)^j s_{j, n+1} \sum_{\rho} A_{n+2}(1, \rho, n{+}1, n{+}2)
\end{aligned}
\end{equation}
This is a new relation between the $n$-point form factor and $2^{n-2}$ $(n+2)$-point scattering amplitudes. We can use this relation to calculate the stringy form factor in the field-theory limit. For $\mathcal{O}=\mathcal{L}_{\phi^{3}}$, the $A_{n+2}$ is the double trace ($(1,2,3,\ldots,n)$ and $(n{+}1, n{+}2)$)scalar amplitude in Yang-Mills-Scalar(YMS) theory with the color ordering $(1, \rho, n{+}1, n{+}2)$. For $\mathcal{O}=\text{Tr}(F^{2})$, the $A_{n+2}$ is the mixed amplitude ($(1,2,\ldots,n)$ are gluons and $(n{+}1,n{+}2)$ are a pairs of scalars) in YMS theroy with the color ordering $(1, \rho, n{+}1, n{+}2)$.

For example, $n=2$, $F^{\mathcal{O}}_2(1,2;q)\propto s_{2,3} A_{4}(1,2,3,4)$. For $\mathcal{O}=\mathcal{L}_{\phi^{3}}$, the integrand is $\text{PT}(1,2)\text{PT}(3,4)$, which corresponds to the double trace amplitude in the YMS theory~\cite{Cachazo:2014xea}. 

\begin{equation}
    F^{\mathcal{L}_{\phi^{3}}}_2(1,2;q)\propto s_{2,3} (\frac{s_{1,3}}{s_{1,2}})=\frac{1}{4}s_{1,2}\,.
\end{equation}
For $\mathcal{O}=\text{Tr}(F^{2})$, the integrand is $(V_{1}V_{2}+W_{1,2})\text{PT}(3,4)$, which corresponds to the $2$ gluons and $2$ scalars mixed amplitude in the YMS theory~\cite{Cachazo:2014xea}, also can be transmuted by the pure Yang-Mills amplitudes~\cite{Cheung:2017ems}.
\begin{equation}
\begin{aligned}
   F^{\text{Tr}(F^{2})}_2&\propto s_{2,3} (\frac{T_{1,2,3}{-}T_{2,1}T_{1,3}}{s_{1,2}}{+}\frac{T_{1,2,3}{+}T_{1,3}T_{2,3}}{s_{2,3}})\\
    &=\frac{1}{8}\text{tr}(1,2)\,.
\end{aligned}
\end{equation}
with the shorthands $T_{1,2,3}=\epsilon _1\cdot f_2\cdot k_3$, and $T_{i,j}=\epsilon _i\cdot k_j$.

We have calculated and checked the relation~\eqref{eq:relation} up to $7$-point, by using the public code in~\cite{He:2021lro} to obtain the data on scattering amplitudes, and compared with the form factors from the pubic code in~\cite{Dong:2022bta}. We believe this non-trivial relations will have deeper understandings in the field-theory, and leave the exploration in the future.

\section{Conclusions and Outlooks}
In this letter, we propose a stringy model as the UV completion of the tree-level form factors from the bosonic open $\&$ closed string disk amplitudes. Here we focus on the two examples: the Lagrangian operator $\mathcal{L}_{\phi^{3}}$ for Tr($\phi^{3}$) theory with the scalar external states, and the Lagrangian operator Tr($F^{2}$) for Yang-Mills theory with the gluon external states. The stringy form factors manifest some hidden properties and uncover the relations with amplitudes. The relations help us to evaluate the form factor in the field-theory limit.

Our investigation of stringy form factors opens up many interesting  directions. A natural extension is to explore super form factors derived from superstring theory in the RNS formalism~\cite{Ramond:1971gb,Neveu:1971rx,Green:1987sp}, or the pure spinor formalism~\cite{Berkovits:2000fe,Berkovits:2004px,Berkovits:2005bt}. Based on our analysis in the bosonic string case, we anticipate that the superstring will yield local operators corresponding to the Lagrangian of Super Yang-Mills theory, the fermion current operator in the R sector or the ${\rm Tr}(F^2)$ operator in the NS sector.

One can also consider inserting additional closed-string vertex operators. With the correct integrand, we expect this to serve as the UV completion of a more general form factor involving multiple local operators, such as $\langle 0| \mathcal{O}(x_1)\mathcal{O}(x_2)\left|1,2, \ldots, n \right\rangle$. Such form factors have been previously studied in the context of $\mathcal{N}=4$ SYM~\cite{Engelund:2012re, Ahmed:2019upm, Ahmed:2019yjt}.

We primarily focus on the field-theory limit of the stringy form factor. However, an interesting direction is to explore higher-order $\alpha'$ corrections. We think the$\alpha^{\prime}$ corrections to stringy form factors have two contributions. One is the correction to the local operator in the form factors, such as higher-dimension operators. The second is the correction to the interactions among the external particles. For instance, examining the next-to-leading order in $\alpha'$ for the ${\rm Tr}(F^2)$ stringy form factor might reveal local operators such as ${\rm Tr}(F^3)$, ${\rm Tr}(F^4)$, and beyond.

Here, we only address the tree-level stringy form factor. An exciting direction for future exploration is extending the concept of stringy form factors to higher-loop cases, which correspond to higher-genus surfaces. For instance, see the discussion on open $\&$ closed strings at one loop~\cite{Stieberger:2021daa}.

Recently, a new program called surfaceology has been widely developed~\cite{Arkani-Hamed:2023lbd,Arkani-Hamed:2023mvg,Arkani-Hamed:2023swr,Arkani-Hamed:2023jry,Arkani-Hamed:2024nhp,Arkani-Hamed:2024vna,Arkani-Hamed:2024fyd,Arkani-Hamed:2024tzl,Arkani-Hamed:2024nzc,Arkani-Hamed:2024pzc,De:2024wsy,Cao:2024olg,Cao:2025mlt}, based on the $u$-variables and binary geometry on the Riemann surfaces~\cite{Arkani-Hamed:2019mrd,Arkani-Hamed:2019plo,Arkani-Hamed:2019vag}. It would be intriguing to explore the connection between our study on stringy form factors and surfaceology. This connection may also provide a more natural path for generalizing to all-loop orders.

Last but not least, an important question is how to apply the double-copy construction or other methods to obtain form factors in gravity theories~\cite{Lin:2021pne,Lin:2022jrp,Lin:2023rwe}. We hope that our approach to form factors in color-ordered theories will provide insights into their connection with gravity form factors.

\section*{acknowledgments}
We thank Jin Dong, Song He , and Fan Zhu for stimulating discussions and collaborations on related projects, and we thank Song He and Yichao Tang for useful comments on the
manuscript. This work is supported by the National Natural Science Foundation of China under Grant No. 123B2075 and No. 12347103.


\bibliographystyle{apsrev4-1}
\bibliography{FF.bib}

\begin{thebibliography}{108}%
\makeatletter
\providecommand \@ifxundefined [1]{%
 \@ifx{#1\undefined}
}%
\providecommand \@ifnum [1]{%
 \ifnum #1\expandafter \@firstoftwo
 \else \expandafter \@secondoftwo
 \fi
}%
\providecommand \@ifx [1]{%
 \ifx #1\expandafter \@firstoftwo
 \else \expandafter \@secondoftwo
 \fi
}%
\providecommand \natexlab [1]{#1}%
\providecommand \enquote  [1]{``#1''}%
\providecommand \bibnamefont  [1]{#1}%
\providecommand \bibfnamefont [1]{#1}%
\providecommand \citenamefont [1]{#1}%
\providecommand \href@noop [0]{\@secondoftwo}%
\providecommand \href [0]{\begingroup \@sanitize@url \@href}%
\providecommand \@href[1]{\@@startlink{#1}\@@href}%
\providecommand \@@href[1]{\endgroup#1\@@endlink}%
\providecommand \@sanitize@url [0]{\catcode `\\12\catcode `\$12\catcode `\&12\catcode `\#12\catcode `\^12\catcode `\_12\catcode `\%12\relax}%
\providecommand \@@startlink[1]{}%
\providecommand \@@endlink[0]{}%
\providecommand \url  [0]{\begingroup\@sanitize@url \@url }%
\providecommand \@url [1]{\endgroup\@href {#1}{\urlprefix }}%
\providecommand \urlprefix  [0]{URL }%
\providecommand \Eprint [0]{\href }%
\providecommand \doibase [0]{http://dx.doi.org/}%
\providecommand \selectlanguage [0]{\@gobble}%
\providecommand \bibinfo  [0]{\@secondoftwo}%
\providecommand \bibfield  [0]{\@secondoftwo}%
\providecommand \translation [1]{[#1]}%
\providecommand \BibitemOpen [0]{}%
\providecommand \bibitemStop [0]{}%
\providecommand \bibitemNoStop [0]{.\EOS\space}%
\providecommand \EOS [0]{\spacefactor3000\relax}%
\providecommand \BibitemShut  [1]{\csname bibitem#1\endcsname}%
\let\auto@bib@innerbib\@empty
\bibitem [{\citenamefont {Kawai}\ \emph {et~al.}(1986)\citenamefont {Kawai}, \citenamefont {Lewellen},\ and\ \citenamefont {Tye}}]{Kawai:1985xq}%
  \BibitemOpen
  \bibfield  {author} {\bibinfo {author} {\bibfnamefont {H.}~\bibnamefont {Kawai}}, \bibinfo {author} {\bibfnamefont {D.~C.}\ \bibnamefont {Lewellen}}, \ and\ \bibinfo {author} {\bibfnamefont {S.~H.~H.}\ \bibnamefont {Tye}},\ }\href {\doibase 10.1016/0550-3213(86)90362-7} {\bibfield  {journal} {\bibinfo  {journal} {Nucl. Phys. B}\ }\textbf {\bibinfo {volume} {269}},\ \bibinfo {pages} {1} (\bibinfo {year} {1986})}\BibitemShut {NoStop}%
\bibitem [{\citenamefont {Bern}\ and\ \citenamefont {Kosower}(1991{\natexlab{a}})}]{Bern:1990cu}%
  \BibitemOpen
  \bibfield  {author} {\bibinfo {author} {\bibfnamefont {Z.}~\bibnamefont {Bern}}\ and\ \bibinfo {author} {\bibfnamefont {D.~A.}\ \bibnamefont {Kosower}},\ }\href {\doibase 10.1103/PhysRevLett.66.1669} {\bibfield  {journal} {\bibinfo  {journal} {Phys. Rev. Lett.}\ }\textbf {\bibinfo {volume} {66}},\ \bibinfo {pages} {1669} (\bibinfo {year} {1991}{\natexlab{a}})}\BibitemShut {NoStop}%
\bibitem [{\citenamefont {Bern}\ and\ \citenamefont {Kosower}(1991{\natexlab{b}})}]{Bern:1990ux}%
  \BibitemOpen
  \bibfield  {author} {\bibinfo {author} {\bibfnamefont {Z.}~\bibnamefont {Bern}}\ and\ \bibinfo {author} {\bibfnamefont {D.~A.}\ \bibnamefont {Kosower}},\ }\href {\doibase 10.1016/0550-3213(91)90567-H} {\bibfield  {journal} {\bibinfo  {journal} {Nucl. Phys. B}\ }\textbf {\bibinfo {volume} {362}},\ \bibinfo {pages} {389} (\bibinfo {year} {1991}{\natexlab{b}})}\BibitemShut {NoStop}%
\bibitem [{\citenamefont {Bern}\ and\ \citenamefont {Kosower}(1992)}]{Bern:1991aq}%
  \BibitemOpen
  \bibfield  {author} {\bibinfo {author} {\bibfnamefont {Z.}~\bibnamefont {Bern}}\ and\ \bibinfo {author} {\bibfnamefont {D.~A.}\ \bibnamefont {Kosower}},\ }\href {\doibase 10.1016/0550-3213(92)90134-W} {\bibfield  {journal} {\bibinfo  {journal} {Nucl. Phys. B}\ }\textbf {\bibinfo {volume} {379}},\ \bibinfo {pages} {451} (\bibinfo {year} {1992})}\BibitemShut {NoStop}%
\bibitem [{\citenamefont {Bern}\ \emph {et~al.}(1998)\citenamefont {Bern}, \citenamefont {Del~Duca},\ and\ \citenamefont {Schmidt}}]{Bern:1998sc}%
  \BibitemOpen
  \bibfield  {author} {\bibinfo {author} {\bibfnamefont {Z.}~\bibnamefont {Bern}}, \bibinfo {author} {\bibfnamefont {V.}~\bibnamefont {Del~Duca}}, \ and\ \bibinfo {author} {\bibfnamefont {C.~R.}\ \bibnamefont {Schmidt}},\ }\href {\doibase 10.1016/S0370-2693(98)01495-6} {\bibfield  {journal} {\bibinfo  {journal} {Phys. Lett. B}\ }\textbf {\bibinfo {volume} {445}},\ \bibinfo {pages} {168} (\bibinfo {year} {1998})},\ \Eprint {http://arxiv.org/abs/hep-ph/9810409} {arXiv:hep-ph/9810409} \BibitemShut {NoStop}%
\bibitem [{\citenamefont {Bern}\ \emph {et~al.}(1999)\citenamefont {Bern}, \citenamefont {Dixon}, \citenamefont {Perelstein},\ and\ \citenamefont {Rozowsky}}]{Bern:1998sv}%
  \BibitemOpen
  \bibfield  {author} {\bibinfo {author} {\bibfnamefont {Z.}~\bibnamefont {Bern}}, \bibinfo {author} {\bibfnamefont {L.~J.}\ \bibnamefont {Dixon}}, \bibinfo {author} {\bibfnamefont {M.}~\bibnamefont {Perelstein}}, \ and\ \bibinfo {author} {\bibfnamefont {J.~S.}\ \bibnamefont {Rozowsky}},\ }\href {\doibase 10.1016/S0550-3213(99)00029-2} {\bibfield  {journal} {\bibinfo  {journal} {Nucl. Phys. B}\ }\textbf {\bibinfo {volume} {546}},\ \bibinfo {pages} {423} (\bibinfo {year} {1999})},\ \Eprint {http://arxiv.org/abs/hep-th/9811140} {arXiv:hep-th/9811140} \BibitemShut {NoStop}%
\bibitem [{\citenamefont {Witten}(2004)}]{Witten:2003nn}%
  \BibitemOpen
  \bibfield  {author} {\bibinfo {author} {\bibfnamefont {E.}~\bibnamefont {Witten}},\ }\href {\doibase 10.1007/s00220-004-1187-3} {\bibfield  {journal} {\bibinfo  {journal} {Commun. Math. Phys.}\ }\textbf {\bibinfo {volume} {252}},\ \bibinfo {pages} {189} (\bibinfo {year} {2004})},\ \Eprint {http://arxiv.org/abs/hep-th/0312171} {arXiv:hep-th/0312171} \BibitemShut {NoStop}%
\bibitem [{\citenamefont {Cachazo}\ \emph {et~al.}(2014{\natexlab{a}})\citenamefont {Cachazo}, \citenamefont {He},\ and\ \citenamefont {Yuan}}]{Cachazo:2013gna}%
  \BibitemOpen
  \bibfield  {author} {\bibinfo {author} {\bibfnamefont {F.}~\bibnamefont {Cachazo}}, \bibinfo {author} {\bibfnamefont {S.}~\bibnamefont {He}}, \ and\ \bibinfo {author} {\bibfnamefont {E.~Y.}\ \bibnamefont {Yuan}},\ }\href {\doibase 10.1103/PhysRevD.90.065001} {\bibfield  {journal} {\bibinfo  {journal} {Phys. Rev. D}\ }\textbf {\bibinfo {volume} {90}},\ \bibinfo {pages} {065001} (\bibinfo {year} {2014}{\natexlab{a}})},\ \Eprint {http://arxiv.org/abs/1306.6575} {arXiv:1306.6575 [hep-th]} \BibitemShut {NoStop}%
\bibitem [{\citenamefont {Cachazo}\ \emph {et~al.}(2014{\natexlab{b}})\citenamefont {Cachazo}, \citenamefont {He},\ and\ \citenamefont {Yuan}}]{Cachazo:2013iea}%
  \BibitemOpen
  \bibfield  {author} {\bibinfo {author} {\bibfnamefont {F.}~\bibnamefont {Cachazo}}, \bibinfo {author} {\bibfnamefont {S.}~\bibnamefont {He}}, \ and\ \bibinfo {author} {\bibfnamefont {E.~Y.}\ \bibnamefont {Yuan}},\ }\href {\doibase 10.1007/JHEP07(2014)033} {\bibfield  {journal} {\bibinfo  {journal} {JHEP}\ }\textbf {\bibinfo {volume} {07}},\ \bibinfo {pages} {033} (\bibinfo {year} {2014}{\natexlab{b}})},\ \Eprint {http://arxiv.org/abs/1309.0885} {arXiv:1309.0885 [hep-th]} \BibitemShut {NoStop}%
\bibitem [{\citenamefont {Cachazo}\ \emph {et~al.}(2014{\natexlab{c}})\citenamefont {Cachazo}, \citenamefont {He},\ and\ \citenamefont {Yuan}}]{Cachazo:2013hca}%
  \BibitemOpen
  \bibfield  {author} {\bibinfo {author} {\bibfnamefont {F.}~\bibnamefont {Cachazo}}, \bibinfo {author} {\bibfnamefont {S.}~\bibnamefont {He}}, \ and\ \bibinfo {author} {\bibfnamefont {E.~Y.}\ \bibnamefont {Yuan}},\ }\href {\doibase 10.1103/PhysRevLett.113.171601} {\bibfield  {journal} {\bibinfo  {journal} {Phys. Rev. Lett.}\ }\textbf {\bibinfo {volume} {113}},\ \bibinfo {pages} {171601} (\bibinfo {year} {2014}{\natexlab{c}})},\ \Eprint {http://arxiv.org/abs/1307.2199} {arXiv:1307.2199 [hep-th]} \BibitemShut {NoStop}%
\bibitem [{\citenamefont {Mason}\ and\ \citenamefont {Skinner}(2014)}]{Mason:2013sva}%
  \BibitemOpen
  \bibfield  {author} {\bibinfo {author} {\bibfnamefont {L.}~\bibnamefont {Mason}}\ and\ \bibinfo {author} {\bibfnamefont {D.}~\bibnamefont {Skinner}},\ }\href {\doibase 10.1007/JHEP07(2014)048} {\bibfield  {journal} {\bibinfo  {journal} {JHEP}\ }\textbf {\bibinfo {volume} {07}},\ \bibinfo {pages} {048} (\bibinfo {year} {2014})},\ \Eprint {http://arxiv.org/abs/1311.2564} {arXiv:1311.2564 [hep-th]} \BibitemShut {NoStop}%
\bibitem [{\citenamefont {He}\ \emph {et~al.}(2019{\natexlab{a}})\citenamefont {He}, \citenamefont {Teng},\ and\ \citenamefont {Zhang}}]{He:2018pol}%
  \BibitemOpen
  \bibfield  {author} {\bibinfo {author} {\bibfnamefont {S.}~\bibnamefont {He}}, \bibinfo {author} {\bibfnamefont {F.}~\bibnamefont {Teng}}, \ and\ \bibinfo {author} {\bibfnamefont {Y.}~\bibnamefont {Zhang}},\ }\href {\doibase 10.1103/PhysRevLett.122.211603} {\bibfield  {journal} {\bibinfo  {journal} {Phys. Rev. Lett.}\ }\textbf {\bibinfo {volume} {122}},\ \bibinfo {pages} {211603} (\bibinfo {year} {2019}{\natexlab{a}})},\ \Eprint {http://arxiv.org/abs/1812.03369} {arXiv:1812.03369 [hep-th]} \BibitemShut {NoStop}%
\bibitem [{\citenamefont {He}\ \emph {et~al.}(2019{\natexlab{b}})\citenamefont {He}, \citenamefont {Teng},\ and\ \citenamefont {Zhang}}]{He:2019drm}%
  \BibitemOpen
  \bibfield  {author} {\bibinfo {author} {\bibfnamefont {S.}~\bibnamefont {He}}, \bibinfo {author} {\bibfnamefont {F.}~\bibnamefont {Teng}}, \ and\ \bibinfo {author} {\bibfnamefont {Y.}~\bibnamefont {Zhang}},\ }\href {\doibase 10.1007/JHEP09(2019)085} {\bibfield  {journal} {\bibinfo  {journal} {JHEP}\ }\textbf {\bibinfo {volume} {09}},\ \bibinfo {pages} {085} (\bibinfo {year} {2019}{\natexlab{b}})},\ \Eprint {http://arxiv.org/abs/1907.06041} {arXiv:1907.06041 [hep-th]} \BibitemShut {NoStop}%
\bibitem [{\citenamefont {Bern}(1992)}]{Bern:1992ad}%
  \BibitemOpen
  \bibfield  {author} {\bibinfo {author} {\bibfnamefont {Z.}~\bibnamefont {Bern}},\ }in\ \href@noop {} {\emph {\bibinfo {booktitle} {{Theoretical Advanced Study Institute (TASI 92): From Black Holes and Strings to Particles}}}}\ (\bibinfo {year} {1992})\ pp.\ \bibinfo {pages} {0471--536},\ \Eprint {http://arxiv.org/abs/hep-ph/9304249} {arXiv:hep-ph/9304249} \BibitemShut {NoStop}%
\bibitem [{\citenamefont {Elvang}\ and\ \citenamefont {Huang}(2013)}]{Elvang:2013cua}%
  \BibitemOpen
  \bibfield  {author} {\bibinfo {author} {\bibfnamefont {H.}~\bibnamefont {Elvang}}\ and\ \bibinfo {author} {\bibfnamefont {Y.-t.}\ \bibnamefont {Huang}},\ }\href@noop {} {\  (\bibinfo {year} {2013})},\ \Eprint {http://arxiv.org/abs/1308.1697} {arXiv:1308.1697 [hep-th]} \BibitemShut {NoStop}%
\bibitem [{\citenamefont {Henn}\ and\ \citenamefont {Plefka}(2014)}]{Henn:2014yza}%
  \BibitemOpen
  \bibfield  {author} {\bibinfo {author} {\bibfnamefont {J.~M.}\ \bibnamefont {Henn}}\ and\ \bibinfo {author} {\bibfnamefont {J.~C.}\ \bibnamefont {Plefka}},\ }\href {\doibase 10.1007/978-3-642-54022-6} {\emph {\bibinfo {title} {{Scattering Amplitudes in Gauge Theories}}}},\ Vol.\ \bibinfo {volume} {883}\ (\bibinfo  {publisher} {Springer},\ \bibinfo {address} {Berlin},\ \bibinfo {year} {2014})\BibitemShut {NoStop}%
\bibitem [{\citenamefont {Travaglini}\ \emph {et~al.}(2022)\citenamefont {Travaglini} \emph {et~al.}}]{Travaglini:2022uwo}%
  \BibitemOpen
  \bibfield  {author} {\bibinfo {author} {\bibfnamefont {G.}~\bibnamefont {Travaglini}} \emph {et~al.},\ }\href {\doibase 10.1088/1751-8121/ac8380} {\bibfield  {journal} {\bibinfo  {journal} {J. Phys. A}\ }\textbf {\bibinfo {volume} {55}},\ \bibinfo {pages} {443001} (\bibinfo {year} {2022})},\ \Eprint {http://arxiv.org/abs/2203.13011} {arXiv:2203.13011 [hep-th]} \BibitemShut {NoStop}%
\bibitem [{\citenamefont {Mafra}\ and\ \citenamefont {Schlotterer}(2023)}]{Mafra:2022wml}%
  \BibitemOpen
  \bibfield  {author} {\bibinfo {author} {\bibfnamefont {C.~R.}\ \bibnamefont {Mafra}}\ and\ \bibinfo {author} {\bibfnamefont {O.}~\bibnamefont {Schlotterer}},\ }\href {\doibase 10.1016/j.physrep.2023.04.001} {\bibfield  {journal} {\bibinfo  {journal} {Phys. Rept.}\ }\textbf {\bibinfo {volume} {1020}},\ \bibinfo {pages} {1} (\bibinfo {year} {2023})},\ \Eprint {http://arxiv.org/abs/2210.14241} {arXiv:2210.14241 [hep-th]} \BibitemShut {NoStop}%
\bibitem [{\citenamefont {Berkovits}\ \emph {et~al.}(2022)\citenamefont {Berkovits}, \citenamefont {D'Hoker}, \citenamefont {Green}, \citenamefont {Johansson},\ and\ \citenamefont {Schlotterer}}]{Berkovits:2022ivl}%
  \BibitemOpen
  \bibfield  {author} {\bibinfo {author} {\bibfnamefont {N.}~\bibnamefont {Berkovits}}, \bibinfo {author} {\bibfnamefont {E.}~\bibnamefont {D'Hoker}}, \bibinfo {author} {\bibfnamefont {M.~B.}\ \bibnamefont {Green}}, \bibinfo {author} {\bibfnamefont {H.}~\bibnamefont {Johansson}}, \ and\ \bibinfo {author} {\bibfnamefont {O.}~\bibnamefont {Schlotterer}},\ }in\ \href@noop {} {\emph {\bibinfo {booktitle} {{Snowmass 2021}}}}\ (\bibinfo {year} {2022})\ \Eprint {http://arxiv.org/abs/2203.09099} {arXiv:2203.09099 [hep-th]} \BibitemShut {NoStop}%
\bibitem [{\citenamefont {Green}\ \emph {et~al.}(1988)\citenamefont {Green}, \citenamefont {Schwarz},\ and\ \citenamefont {Witten}}]{Green:1987sp}%
  \BibitemOpen
  \bibfield  {author} {\bibinfo {author} {\bibfnamefont {M.~B.}\ \bibnamefont {Green}}, \bibinfo {author} {\bibfnamefont {J.~H.}\ \bibnamefont {Schwarz}}, \ and\ \bibinfo {author} {\bibfnamefont {E.}~\bibnamefont {Witten}},\ }\href@noop {} {\emph {\bibinfo {title} {{SUPERSTRING THEORY. VOL. 1: INTRODUCTION}}}},\ Cambridge Monographs on Mathematical Physics\ (\bibinfo {year} {1988})\BibitemShut {NoStop}%
\bibitem [{\citenamefont {Stieberger}(2014)}]{Stieberger:2013wea}%
  \BibitemOpen
  \bibfield  {author} {\bibinfo {author} {\bibfnamefont {S.}~\bibnamefont {Stieberger}},\ }\href {\doibase 10.1088/1751-8113/47/15/155401} {\bibfield  {journal} {\bibinfo  {journal} {J. Phys. A}\ }\textbf {\bibinfo {volume} {47}},\ \bibinfo {pages} {155401} (\bibinfo {year} {2014})},\ \Eprint {http://arxiv.org/abs/1310.3259} {arXiv:1310.3259 [hep-th]} \BibitemShut {NoStop}%
\bibitem [{\citenamefont {Stieberger}\ and\ \citenamefont {Taylor}(2014)}]{Stieberger:2014hba}%
  \BibitemOpen
  \bibfield  {author} {\bibinfo {author} {\bibfnamefont {S.}~\bibnamefont {Stieberger}}\ and\ \bibinfo {author} {\bibfnamefont {T.~R.}\ \bibnamefont {Taylor}},\ }\href {\doibase 10.1016/j.nuclphysb.2014.02.005} {\bibfield  {journal} {\bibinfo  {journal} {Nucl. Phys. B}\ }\textbf {\bibinfo {volume} {881}},\ \bibinfo {pages} {269} (\bibinfo {year} {2014})},\ \Eprint {http://arxiv.org/abs/1401.1218} {arXiv:1401.1218 [hep-th]} \BibitemShut {NoStop}%
\bibitem [{\citenamefont {Arkani-Hamed}\ \emph {et~al.}(2021)\citenamefont {Arkani-Hamed}, \citenamefont {He},\ and\ \citenamefont {Lam}}]{Arkani-Hamed:2019mrd}%
  \BibitemOpen
  \bibfield  {author} {\bibinfo {author} {\bibfnamefont {N.}~\bibnamefont {Arkani-Hamed}}, \bibinfo {author} {\bibfnamefont {S.}~\bibnamefont {He}}, \ and\ \bibinfo {author} {\bibfnamefont {T.}~\bibnamefont {Lam}},\ }\href {\doibase 10.1007/JHEP02(2021)069} {\bibfield  {journal} {\bibinfo  {journal} {JHEP}\ }\textbf {\bibinfo {volume} {02}},\ \bibinfo {pages} {069} (\bibinfo {year} {2021})},\ \Eprint {http://arxiv.org/abs/1912.08707} {arXiv:1912.08707 [hep-th]} \BibitemShut {NoStop}%
\bibitem [{\citenamefont {Bjerrum-Bohr}\ \emph {et~al.}(2009)\citenamefont {Bjerrum-Bohr}, \citenamefont {Damgaard},\ and\ \citenamefont {Vanhove}}]{Bjerrum-Bohr:2009ulz}%
  \BibitemOpen
  \bibfield  {author} {\bibinfo {author} {\bibfnamefont {N.~E.~J.}\ \bibnamefont {Bjerrum-Bohr}}, \bibinfo {author} {\bibfnamefont {P.~H.}\ \bibnamefont {Damgaard}}, \ and\ \bibinfo {author} {\bibfnamefont {P.}~\bibnamefont {Vanhove}},\ }\href {\doibase 10.1103/PhysRevLett.103.161602} {\bibfield  {journal} {\bibinfo  {journal} {Phys. Rev. Lett.}\ }\textbf {\bibinfo {volume} {103}},\ \bibinfo {pages} {161602} (\bibinfo {year} {2009})},\ \Eprint {http://arxiv.org/abs/0907.1425} {arXiv:0907.1425 [hep-th]} \BibitemShut {NoStop}%
\bibitem [{\citenamefont {Stieberger}(2009)}]{Stieberger:2009hq}%
  \BibitemOpen
  \bibfield  {author} {\bibinfo {author} {\bibfnamefont {S.}~\bibnamefont {Stieberger}},\ }\href@noop {} {\  (\bibinfo {year} {2009})},\ \Eprint {http://arxiv.org/abs/0907.2211} {arXiv:0907.2211 [hep-th]} \BibitemShut {NoStop}%
\bibitem [{\citenamefont {Kleiss}\ and\ \citenamefont {Kuijf}(1989)}]{Kleiss:1988ne}%
  \BibitemOpen
  \bibfield  {author} {\bibinfo {author} {\bibfnamefont {R.}~\bibnamefont {Kleiss}}\ and\ \bibinfo {author} {\bibfnamefont {H.}~\bibnamefont {Kuijf}},\ }\href {\doibase 10.1016/0550-3213(89)90574-9} {\bibfield  {journal} {\bibinfo  {journal} {Nucl. Phys. B}\ }\textbf {\bibinfo {volume} {312}},\ \bibinfo {pages} {616} (\bibinfo {year} {1989})}\BibitemShut {NoStop}%
\bibitem [{\citenamefont {Bern}\ \emph {et~al.}(2008)\citenamefont {Bern}, \citenamefont {Carrasco},\ and\ \citenamefont {Johansson}}]{Bern:2008qj}%
  \BibitemOpen
  \bibfield  {author} {\bibinfo {author} {\bibfnamefont {Z.}~\bibnamefont {Bern}}, \bibinfo {author} {\bibfnamefont {J.~J.~M.}\ \bibnamefont {Carrasco}}, \ and\ \bibinfo {author} {\bibfnamefont {H.}~\bibnamefont {Johansson}},\ }\href {\doibase 10.1103/PhysRevD.78.085011} {\bibfield  {journal} {\bibinfo  {journal} {Phys. Rev. D}\ }\textbf {\bibinfo {volume} {78}},\ \bibinfo {pages} {085011} (\bibinfo {year} {2008})},\ \Eprint {http://arxiv.org/abs/0805.3993} {arXiv:0805.3993 [hep-ph]} \BibitemShut {NoStop}%
\bibitem [{\citenamefont {van Neerven}(1986)}]{vanNeerven:1985ja}%
  \BibitemOpen
  \bibfield  {author} {\bibinfo {author} {\bibfnamefont {W.~L.}\ \bibnamefont {van Neerven}},\ }\href {\doibase 10.1007/BF01571808} {\bibfield  {journal} {\bibinfo  {journal} {Z. Phys. C}\ }\textbf {\bibinfo {volume} {30}},\ \bibinfo {pages} {595} (\bibinfo {year} {1986})}\BibitemShut {NoStop}%
\bibitem [{\citenamefont {Maldacena}\ and\ \citenamefont {Zhiboedov}(2010)}]{Maldacena:2010kp}%
  \BibitemOpen
  \bibfield  {author} {\bibinfo {author} {\bibfnamefont {J.}~\bibnamefont {Maldacena}}\ and\ \bibinfo {author} {\bibfnamefont {A.}~\bibnamefont {Zhiboedov}},\ }\href {\doibase 10.1007/JHEP11(2010)104} {\bibfield  {journal} {\bibinfo  {journal} {JHEP}\ }\textbf {\bibinfo {volume} {11}},\ \bibinfo {pages} {104} (\bibinfo {year} {2010})},\ \Eprint {http://arxiv.org/abs/1009.1139} {arXiv:1009.1139 [hep-th]} \BibitemShut {NoStop}%
\bibitem [{\citenamefont {Brandhuber}\ \emph {et~al.}(2011{\natexlab{a}})\citenamefont {Brandhuber}, \citenamefont {Spence}, \citenamefont {Travaglini},\ and\ \citenamefont {Yang}}]{Brandhuber:2010ad}%
  \BibitemOpen
  \bibfield  {author} {\bibinfo {author} {\bibfnamefont {A.}~\bibnamefont {Brandhuber}}, \bibinfo {author} {\bibfnamefont {B.}~\bibnamefont {Spence}}, \bibinfo {author} {\bibfnamefont {G.}~\bibnamefont {Travaglini}}, \ and\ \bibinfo {author} {\bibfnamefont {G.}~\bibnamefont {Yang}},\ }\href {\doibase 10.1007/JHEP01(2011)134} {\bibfield  {journal} {\bibinfo  {journal} {JHEP}\ }\textbf {\bibinfo {volume} {01}},\ \bibinfo {pages} {134} (\bibinfo {year} {2011}{\natexlab{a}})},\ \Eprint {http://arxiv.org/abs/1011.1899} {arXiv:1011.1899 [hep-th]} \BibitemShut {NoStop}%
\bibitem [{\citenamefont {Bork}\ \emph {et~al.}(2011)\citenamefont {Bork}, \citenamefont {Kazakov},\ and\ \citenamefont {Vartanov}}]{Bork:2010wf}%
  \BibitemOpen
  \bibfield  {author} {\bibinfo {author} {\bibfnamefont {L.~V.}\ \bibnamefont {Bork}}, \bibinfo {author} {\bibfnamefont {D.~I.}\ \bibnamefont {Kazakov}}, \ and\ \bibinfo {author} {\bibfnamefont {G.~S.}\ \bibnamefont {Vartanov}},\ }\href {\doibase 10.1007/JHEP02(2011)063} {\bibfield  {journal} {\bibinfo  {journal} {JHEP}\ }\textbf {\bibinfo {volume} {02}},\ \bibinfo {pages} {063} (\bibinfo {year} {2011})},\ \Eprint {http://arxiv.org/abs/1011.2440} {arXiv:1011.2440 [hep-th]} \BibitemShut {NoStop}%
\bibitem [{\citenamefont {Brandhuber}\ \emph {et~al.}(2011{\natexlab{b}})\citenamefont {Brandhuber}, \citenamefont {Gurdogan}, \citenamefont {Mooney}, \citenamefont {Travaglini},\ and\ \citenamefont {Yang}}]{Brandhuber:2011tv}%
  \BibitemOpen
  \bibfield  {author} {\bibinfo {author} {\bibfnamefont {A.}~\bibnamefont {Brandhuber}}, \bibinfo {author} {\bibfnamefont {O.}~\bibnamefont {Gurdogan}}, \bibinfo {author} {\bibfnamefont {R.}~\bibnamefont {Mooney}}, \bibinfo {author} {\bibfnamefont {G.}~\bibnamefont {Travaglini}}, \ and\ \bibinfo {author} {\bibfnamefont {G.}~\bibnamefont {Yang}},\ }\href {\doibase 10.1007/JHEP10(2011)046} {\bibfield  {journal} {\bibinfo  {journal} {JHEP}\ }\textbf {\bibinfo {volume} {10}},\ \bibinfo {pages} {046} (\bibinfo {year} {2011}{\natexlab{b}})},\ \Eprint {http://arxiv.org/abs/1107.5067} {arXiv:1107.5067 [hep-th]} \BibitemShut {NoStop}%
\bibitem [{\citenamefont {Brandhuber}\ \emph {et~al.}(2012)\citenamefont {Brandhuber}, \citenamefont {Travaglini},\ and\ \citenamefont {Yang}}]{Brandhuber:2012vm}%
  \BibitemOpen
  \bibfield  {author} {\bibinfo {author} {\bibfnamefont {A.}~\bibnamefont {Brandhuber}}, \bibinfo {author} {\bibfnamefont {G.}~\bibnamefont {Travaglini}}, \ and\ \bibinfo {author} {\bibfnamefont {G.}~\bibnamefont {Yang}},\ }\href {\doibase 10.1007/JHEP05(2012)082} {\bibfield  {journal} {\bibinfo  {journal} {JHEP}\ }\textbf {\bibinfo {volume} {05}},\ \bibinfo {pages} {082} (\bibinfo {year} {2012})},\ \Eprint {http://arxiv.org/abs/1201.4170} {arXiv:1201.4170 [hep-th]} \BibitemShut {NoStop}%
\bibitem [{\citenamefont {Brandhuber}\ \emph {et~al.}(2014)\citenamefont {Brandhuber}, \citenamefont {Penante}, \citenamefont {Travaglini},\ and\ \citenamefont {Wen}}]{Brandhuber:2014ica}%
  \BibitemOpen
  \bibfield  {author} {\bibinfo {author} {\bibfnamefont {A.}~\bibnamefont {Brandhuber}}, \bibinfo {author} {\bibfnamefont {B.}~\bibnamefont {Penante}}, \bibinfo {author} {\bibfnamefont {G.}~\bibnamefont {Travaglini}}, \ and\ \bibinfo {author} {\bibfnamefont {C.}~\bibnamefont {Wen}},\ }\href {\doibase 10.1007/JHEP08(2014)100} {\bibfield  {journal} {\bibinfo  {journal} {JHEP}\ }\textbf {\bibinfo {volume} {08}},\ \bibinfo {pages} {100} (\bibinfo {year} {2014})},\ \Eprint {http://arxiv.org/abs/1406.1443} {arXiv:1406.1443 [hep-th]} \BibitemShut {NoStop}%
\bibitem [{\citenamefont {Boels}\ \emph {et~al.}(2013)\citenamefont {Boels}, \citenamefont {Kniehl}, \citenamefont {Tarasov},\ and\ \citenamefont {Yang}}]{Boels:2012ew}%
  \BibitemOpen
  \bibfield  {author} {\bibinfo {author} {\bibfnamefont {R.~H.}\ \bibnamefont {Boels}}, \bibinfo {author} {\bibfnamefont {B.~A.}\ \bibnamefont {Kniehl}}, \bibinfo {author} {\bibfnamefont {O.~V.}\ \bibnamefont {Tarasov}}, \ and\ \bibinfo {author} {\bibfnamefont {G.}~\bibnamefont {Yang}},\ }\href {\doibase 10.1007/JHEP02(2013)063} {\bibfield  {journal} {\bibinfo  {journal} {JHEP}\ }\textbf {\bibinfo {volume} {02}},\ \bibinfo {pages} {063} (\bibinfo {year} {2013})},\ \Eprint {http://arxiv.org/abs/1211.7028} {arXiv:1211.7028 [hep-th]} \BibitemShut {NoStop}%
\bibitem [{\citenamefont {Bork}(2014)}]{Bork:2014eqa}%
  \BibitemOpen
  \bibfield  {author} {\bibinfo {author} {\bibfnamefont {L.~V.}\ \bibnamefont {Bork}},\ }\href {\doibase 10.1007/JHEP12(2014)111} {\bibfield  {journal} {\bibinfo  {journal} {JHEP}\ }\textbf {\bibinfo {volume} {12}},\ \bibinfo {pages} {111} (\bibinfo {year} {2014})},\ \Eprint {http://arxiv.org/abs/1407.5568} {arXiv:1407.5568 [hep-th]} \BibitemShut {NoStop}%
\bibitem [{\citenamefont {Frassek}\ \emph {et~al.}(2016)\citenamefont {Frassek}, \citenamefont {Meidinger}, \citenamefont {Nandan},\ and\ \citenamefont {Wilhelm}}]{Frassek:2015rka}%
  \BibitemOpen
  \bibfield  {author} {\bibinfo {author} {\bibfnamefont {R.}~\bibnamefont {Frassek}}, \bibinfo {author} {\bibfnamefont {D.}~\bibnamefont {Meidinger}}, \bibinfo {author} {\bibfnamefont {D.}~\bibnamefont {Nandan}}, \ and\ \bibinfo {author} {\bibfnamefont {M.}~\bibnamefont {Wilhelm}},\ }\href {\doibase 10.1007/JHEP01(2016)182} {\bibfield  {journal} {\bibinfo  {journal} {JHEP}\ }\textbf {\bibinfo {volume} {01}},\ \bibinfo {pages} {182} (\bibinfo {year} {2016})},\ \Eprint {http://arxiv.org/abs/1506.08192} {arXiv:1506.08192 [hep-th]} \BibitemShut {NoStop}%
\bibitem [{\citenamefont {Bork}\ and\ \citenamefont {Onishchenko}(2016)}]{Bork:2016hst}%
  \BibitemOpen
  \bibfield  {author} {\bibinfo {author} {\bibfnamefont {L.~V.}\ \bibnamefont {Bork}}\ and\ \bibinfo {author} {\bibfnamefont {A.~I.}\ \bibnamefont {Onishchenko}},\ }\href {\doibase 10.1007/JHEP12(2016)076} {\bibfield  {journal} {\bibinfo  {journal} {JHEP}\ }\textbf {\bibinfo {volume} {12}},\ \bibinfo {pages} {076} (\bibinfo {year} {2016})},\ \Eprint {http://arxiv.org/abs/1607.00503} {arXiv:1607.00503 [hep-th]} \BibitemShut {NoStop}%
\bibitem [{\citenamefont {Bork}\ and\ \citenamefont {Onishchenko}(2017)}]{Bork:2016xfn}%
  \BibitemOpen
  \bibfield  {author} {\bibinfo {author} {\bibfnamefont {L.~V.}\ \bibnamefont {Bork}}\ and\ \bibinfo {author} {\bibfnamefont {A.~I.}\ \bibnamefont {Onishchenko}},\ }\href {\doibase 10.1007/JHEP04(2017)019} {\bibfield  {journal} {\bibinfo  {journal} {JHEP}\ }\textbf {\bibinfo {volume} {04}},\ \bibinfo {pages} {019} (\bibinfo {year} {2017})},\ \Eprint {http://arxiv.org/abs/1607.02320} {arXiv:1607.02320 [hep-th]} \BibitemShut {NoStop}%
\bibitem [{\citenamefont {Yang}(2016)}]{Yang:2016ear}%
  \BibitemOpen
  \bibfield  {author} {\bibinfo {author} {\bibfnamefont {G.}~\bibnamefont {Yang}},\ }\href {\doibase 10.1103/PhysRevLett.117.271602} {\bibfield  {journal} {\bibinfo  {journal} {Phys. Rev. Lett.}\ }\textbf {\bibinfo {volume} {117}},\ \bibinfo {pages} {271602} (\bibinfo {year} {2016})},\ \Eprint {http://arxiv.org/abs/1610.02394} {arXiv:1610.02394 [hep-th]} \BibitemShut {NoStop}%
\bibitem [{\citenamefont {Bork}\ and\ \citenamefont {Onishchenko}(2018)}]{Bork:2017qyh}%
  \BibitemOpen
  \bibfield  {author} {\bibinfo {author} {\bibfnamefont {L.~V.}\ \bibnamefont {Bork}}\ and\ \bibinfo {author} {\bibfnamefont {A.~I.}\ \bibnamefont {Onishchenko}},\ }\href {\doibase 10.1103/PhysRevD.97.126013} {\bibfield  {journal} {\bibinfo  {journal} {Phys. Rev. D}\ }\textbf {\bibinfo {volume} {97}},\ \bibinfo {pages} {126013} (\bibinfo {year} {2018})},\ \Eprint {http://arxiv.org/abs/1704.04758} {arXiv:1704.04758 [hep-th]} \BibitemShut {NoStop}%
\bibitem [{\citenamefont {Bianchi}\ \emph {et~al.}(2019)\citenamefont {Bianchi}, \citenamefont {Brandhuber}, \citenamefont {Panerai},\ and\ \citenamefont {Travaglini}}]{Bianchi:2018peu}%
  \BibitemOpen
  \bibfield  {author} {\bibinfo {author} {\bibfnamefont {L.}~\bibnamefont {Bianchi}}, \bibinfo {author} {\bibfnamefont {A.}~\bibnamefont {Brandhuber}}, \bibinfo {author} {\bibfnamefont {R.}~\bibnamefont {Panerai}}, \ and\ \bibinfo {author} {\bibfnamefont {G.}~\bibnamefont {Travaglini}},\ }\href {\doibase 10.1007/JHEP02(2019)182} {\bibfield  {journal} {\bibinfo  {journal} {JHEP}\ }\textbf {\bibinfo {volume} {02}},\ \bibinfo {pages} {182} (\bibinfo {year} {2019})},\ \Eprint {http://arxiv.org/abs/1812.09001} {arXiv:1812.09001 [hep-th]} \BibitemShut {NoStop}%
\bibitem [{\citenamefont {Nandan}\ and\ \citenamefont {Yang}(2018)}]{Nandan:2018hqz}%
  \BibitemOpen
  \bibfield  {author} {\bibinfo {author} {\bibfnamefont {D.}~\bibnamefont {Nandan}}\ and\ \bibinfo {author} {\bibfnamefont {G.}~\bibnamefont {Yang}},\ }\enquote {\bibinfo {title} {{Hidden structure in the form factors of N = 4 SYM}},}\ \ (\bibinfo {year} {2018})\BibitemShut {NoStop}%
\bibitem [{\citenamefont {Yang}(2020)}]{Yang:2019vag}%
  \BibitemOpen
  \bibfield  {author} {\bibinfo {author} {\bibfnamefont {G.}~\bibnamefont {Yang}},\ }\href {\doibase 10.1007/s11433-019-1507-0} {\bibfield  {journal} {\bibinfo  {journal} {Sci. China Phys. Mech. Astron.}\ }\textbf {\bibinfo {volume} {63}},\ \bibinfo {pages} {270001} (\bibinfo {year} {2020})},\ \Eprint {http://arxiv.org/abs/1912.11454} {arXiv:1912.11454 [hep-th]} \BibitemShut {NoStop}%
\bibitem [{\citenamefont {Lin}\ and\ \citenamefont {Yang}(2024{\natexlab{a}})}]{Lin:2022jrp}%
  \BibitemOpen
  \bibfield  {author} {\bibinfo {author} {\bibfnamefont {G.}~\bibnamefont {Lin}}\ and\ \bibinfo {author} {\bibfnamefont {G.}~\bibnamefont {Yang}},\ }\href {\doibase 10.1007/JHEP02(2024)012} {\bibfield  {journal} {\bibinfo  {journal} {JHEP}\ }\textbf {\bibinfo {volume} {02}},\ \bibinfo {pages} {012} (\bibinfo {year} {2024}{\natexlab{a}})},\ \Eprint {http://arxiv.org/abs/2211.01386} {arXiv:2211.01386 [hep-th]} \BibitemShut {NoStop}%
\bibitem [{\citenamefont {Lin}\ and\ \citenamefont {Yang}(2024{\natexlab{b}})}]{Lin:2023rwe}%
  \BibitemOpen
  \bibfield  {author} {\bibinfo {author} {\bibfnamefont {G.}~\bibnamefont {Lin}}\ and\ \bibinfo {author} {\bibfnamefont {G.}~\bibnamefont {Yang}},\ }\href {\doibase 10.1007/JHEP02(2024)013} {\bibfield  {journal} {\bibinfo  {journal} {JHEP}\ }\textbf {\bibinfo {volume} {02}},\ \bibinfo {pages} {013} (\bibinfo {year} {2024}{\natexlab{b}})},\ \Eprint {http://arxiv.org/abs/2306.04672} {arXiv:2306.04672 [hep-th]} \BibitemShut {NoStop}%
\bibitem [{Note1()}]{Note1}%
  \BibitemOpen
  \bibinfo {note} {To be precise, while we believe that a stringy model of form factors may exist on higher-genus Riemann surfaces, we restrict our focus here to the simplest case: the disk topology, which corresponds to the tree-level.}\BibitemShut {Stop}%
\bibitem [{Note2()}]{Note2}%
  \BibitemOpen
  \bibinfo {note} {Some literature studies the one massive string amplitudes~\cite {Guillen:2021mwp,Kashyap:2023cdi,Kashyap:2024qor,Mafra:2024fiy}.}\BibitemShut {Stop}%
\bibitem [{\citenamefont {Klebanov}\ and\ \citenamefont {Thorlacius}(1996)}]{Klebanov:1995ni}%
  \BibitemOpen
  \bibfield  {author} {\bibinfo {author} {\bibfnamefont {I.~R.}\ \bibnamefont {Klebanov}}\ and\ \bibinfo {author} {\bibfnamefont {L.}~\bibnamefont {Thorlacius}},\ }\href {\doibase 10.1016/0370-2693(95)01576-0} {\bibfield  {journal} {\bibinfo  {journal} {Phys. Lett. B}\ }\textbf {\bibinfo {volume} {371}},\ \bibinfo {pages} {51} (\bibinfo {year} {1996})},\ \Eprint {http://arxiv.org/abs/hep-th/9510200} {arXiv:hep-th/9510200} \BibitemShut {NoStop}%
\bibitem [{\citenamefont {Gubser}\ \emph {et~al.}(1996)\citenamefont {Gubser}, \citenamefont {Hashimoto}, \citenamefont {Klebanov},\ and\ \citenamefont {Maldacena}}]{Gubser:1996wt}%
  \BibitemOpen
  \bibfield  {author} {\bibinfo {author} {\bibfnamefont {S.~S.}\ \bibnamefont {Gubser}}, \bibinfo {author} {\bibfnamefont {A.}~\bibnamefont {Hashimoto}}, \bibinfo {author} {\bibfnamefont {I.~R.}\ \bibnamefont {Klebanov}}, \ and\ \bibinfo {author} {\bibfnamefont {J.~M.}\ \bibnamefont {Maldacena}},\ }\href {\doibase 10.1016/0550-3213(96)00182-4} {\bibfield  {journal} {\bibinfo  {journal} {Nucl. Phys. B}\ }\textbf {\bibinfo {volume} {472}},\ \bibinfo {pages} {231} (\bibinfo {year} {1996})},\ \Eprint {http://arxiv.org/abs/hep-th/9601057} {arXiv:hep-th/9601057} \BibitemShut {NoStop}%
\bibitem [{\citenamefont {Garousi}\ and\ \citenamefont {Myers}(1996)}]{Garousi:1996ad}%
  \BibitemOpen
  \bibfield  {author} {\bibinfo {author} {\bibfnamefont {M.~R.}\ \bibnamefont {Garousi}}\ and\ \bibinfo {author} {\bibfnamefont {R.~C.}\ \bibnamefont {Myers}},\ }\href {\doibase 10.1016/0550-3213(96)00316-1} {\bibfield  {journal} {\bibinfo  {journal} {Nucl. Phys. B}\ }\textbf {\bibinfo {volume} {475}},\ \bibinfo {pages} {193} (\bibinfo {year} {1996})},\ \Eprint {http://arxiv.org/abs/hep-th/9603194} {arXiv:hep-th/9603194} \BibitemShut {NoStop}%
\bibitem [{\citenamefont {Hashimoto}\ and\ \citenamefont {Klebanov}(1997)}]{Hashimoto:1996bf}%
  \BibitemOpen
  \bibfield  {author} {\bibinfo {author} {\bibfnamefont {A.}~\bibnamefont {Hashimoto}}\ and\ \bibinfo {author} {\bibfnamefont {I.~R.}\ \bibnamefont {Klebanov}},\ }\href {\doibase 10.1016/S0920-5632(97)00074-1} {\bibfield  {journal} {\bibinfo  {journal} {Nucl. Phys. B Proc. Suppl.}\ }\textbf {\bibinfo {volume} {55}},\ \bibinfo {pages} {118} (\bibinfo {year} {1997})},\ \Eprint {http://arxiv.org/abs/hep-th/9611214} {arXiv:hep-th/9611214} \BibitemShut {NoStop}%
\bibitem [{\citenamefont {Hashimoto}\ and\ \citenamefont {Klebanov}(1996)}]{Hashimoto:1996kf}%
  \BibitemOpen
  \bibfield  {author} {\bibinfo {author} {\bibfnamefont {A.}~\bibnamefont {Hashimoto}}\ and\ \bibinfo {author} {\bibfnamefont {I.~R.}\ \bibnamefont {Klebanov}},\ }\href {\doibase 10.1016/0370-2693(96)00621-1} {\bibfield  {journal} {\bibinfo  {journal} {Phys. Lett. B}\ }\textbf {\bibinfo {volume} {381}},\ \bibinfo {pages} {437} (\bibinfo {year} {1996})},\ \Eprint {http://arxiv.org/abs/hep-th/9604065} {arXiv:hep-th/9604065} \BibitemShut {NoStop}%
\bibitem [{\citenamefont {Sen}(2003{\natexlab{a}})}]{Sen:2003bc}%
  \BibitemOpen
  \bibfield  {author} {\bibinfo {author} {\bibfnamefont {A.}~\bibnamefont {Sen}},\ }\href {\doibase 10.1103/PhysRevD.68.106003} {\bibfield  {journal} {\bibinfo  {journal} {Phys. Rev. D}\ }\textbf {\bibinfo {volume} {68}},\ \bibinfo {pages} {106003} (\bibinfo {year} {2003}{\natexlab{a}})},\ \Eprint {http://arxiv.org/abs/hep-th/0305011} {arXiv:hep-th/0305011} \BibitemShut {NoStop}%
\bibitem [{\citenamefont {Sen}(2003{\natexlab{b}})}]{Sen:2003xs}%
  \BibitemOpen
  \bibfield  {author} {\bibinfo {author} {\bibfnamefont {A.}~\bibnamefont {Sen}},\ }\href {\doibase 10.1103/PhysRevLett.91.181601} {\bibfield  {journal} {\bibinfo  {journal} {Phys. Rev. Lett.}\ }\textbf {\bibinfo {volume} {91}},\ \bibinfo {pages} {181601} (\bibinfo {year} {2003}{\natexlab{b}})},\ \Eprint {http://arxiv.org/abs/hep-th/0306137} {arXiv:hep-th/0306137} \BibitemShut {NoStop}%
\bibitem [{\citenamefont {Stieberger}\ and\ \citenamefont {Taylor}(2016)}]{Stieberger:2015vya}%
  \BibitemOpen
  \bibfield  {author} {\bibinfo {author} {\bibfnamefont {S.}~\bibnamefont {Stieberger}}\ and\ \bibinfo {author} {\bibfnamefont {T.~R.}\ \bibnamefont {Taylor}},\ }\href {\doibase 10.1016/j.nuclphysb.2015.12.002} {\bibfield  {journal} {\bibinfo  {journal} {Nucl. Phys. B}\ }\textbf {\bibinfo {volume} {903}},\ \bibinfo {pages} {104} (\bibinfo {year} {2016})},\ \Eprint {http://arxiv.org/abs/1510.01774} {arXiv:1510.01774 [hep-th]} \BibitemShut {NoStop}%
\bibitem [{Note3()}]{Note3}%
  \BibitemOpen
  \bibinfo {note} {If we actually choose the metric matrix $D^{\mu \nu }$ (the same notation as in~\cite {Stieberger:2015vya}) to be a specific value, we cannot realize the kinematic condition for the form factor ($q^2 \protect \neq 0$). Here we actually generalize the kinematics part of disk string amplitude to the stringy form factor in~\protect \eqref {eq:stringy ff def} by mapping the kinematics $2q^{1} \cdot q^{2} \to q^2, 2p_i \cdot q^{1} = 2p_i \cdot q^{2} \to -p_i \cdot q$, ($q^{1}, q^{2}$ are defined in~\cite {Stieberger:2015vya}). \par This map cannot be realized by any choice of the matrix $D$.}\BibitemShut {Stop}%
\bibitem [{\citenamefont {Koba}\ and\ \citenamefont {Nielsen}(1969)}]{Koba:1969rw}%
  \BibitemOpen
  \bibfield  {author} {\bibinfo {author} {\bibfnamefont {Z.}~\bibnamefont {Koba}}\ and\ \bibinfo {author} {\bibfnamefont {H.~B.}\ \bibnamefont {Nielsen}},\ }\href {\doibase 10.1016/0550-3213(69)90331-9} {\bibfield  {journal} {\bibinfo  {journal} {Nucl. Phys. B}\ }\textbf {\bibinfo {volume} {10}},\ \bibinfo {pages} {633} (\bibinfo {year} {1969})}\BibitemShut {NoStop}%
\bibitem [{\citenamefont {Parke}\ and\ \citenamefont {Taylor}(1986)}]{Parke:1986gb}%
  \BibitemOpen
  \bibfield  {author} {\bibinfo {author} {\bibfnamefont {S.~J.}\ \bibnamefont {Parke}}\ and\ \bibinfo {author} {\bibfnamefont {T.~R.}\ \bibnamefont {Taylor}},\ }\href {\doibase 10.1103/PhysRevLett.56.2459} {\bibfield  {journal} {\bibinfo  {journal} {Phys. Rev. Lett.}\ }\textbf {\bibinfo {volume} {56}},\ \bibinfo {pages} {2459} (\bibinfo {year} {1986})}\BibitemShut {NoStop}%
\bibitem [{\citenamefont {Broedel}\ \emph {et~al.}(2013)\citenamefont {Broedel}, \citenamefont {Schlotterer},\ and\ \citenamefont {Stieberger}}]{Broedel:2013tta}%
  \BibitemOpen
  \bibfield  {author} {\bibinfo {author} {\bibfnamefont {J.}~\bibnamefont {Broedel}}, \bibinfo {author} {\bibfnamefont {O.}~\bibnamefont {Schlotterer}}, \ and\ \bibinfo {author} {\bibfnamefont {S.}~\bibnamefont {Stieberger}},\ }\href {\doibase 10.1002/prop.201300019} {\bibfield  {journal} {\bibinfo  {journal} {Fortsch. Phys.}\ }\textbf {\bibinfo {volume} {61}},\ \bibinfo {pages} {812} (\bibinfo {year} {2013})},\ \Eprint {http://arxiv.org/abs/1304.7267} {arXiv:1304.7267 [hep-th]} \BibitemShut {NoStop}%
\bibitem [{\citenamefont {Carrasco}\ \emph {et~al.}(2017{\natexlab{a}})\citenamefont {Carrasco}, \citenamefont {Mafra},\ and\ \citenamefont {Schlotterer}}]{Carrasco:2016ldy}%
  \BibitemOpen
  \bibfield  {author} {\bibinfo {author} {\bibfnamefont {J.~J.~M.}\ \bibnamefont {Carrasco}}, \bibinfo {author} {\bibfnamefont {C.~R.}\ \bibnamefont {Mafra}}, \ and\ \bibinfo {author} {\bibfnamefont {O.}~\bibnamefont {Schlotterer}},\ }\href {\doibase 10.1007/JHEP06(2017)093} {\bibfield  {journal} {\bibinfo  {journal} {JHEP}\ }\textbf {\bibinfo {volume} {06}},\ \bibinfo {pages} {093} (\bibinfo {year} {2017}{\natexlab{a}})},\ \Eprint {http://arxiv.org/abs/1608.02569} {arXiv:1608.02569 [hep-th]} \BibitemShut {NoStop}%
\bibitem [{\citenamefont {Carrasco}\ \emph {et~al.}(2017{\natexlab{b}})\citenamefont {Carrasco}, \citenamefont {Mafra},\ and\ \citenamefont {Schlotterer}}]{Carrasco:2016ygv}%
  \BibitemOpen
  \bibfield  {author} {\bibinfo {author} {\bibfnamefont {J.~J.~M.}\ \bibnamefont {Carrasco}}, \bibinfo {author} {\bibfnamefont {C.~R.}\ \bibnamefont {Mafra}}, \ and\ \bibinfo {author} {\bibfnamefont {O.}~\bibnamefont {Schlotterer}},\ }\href {\doibase 10.1007/JHEP08(2017)135} {\bibfield  {journal} {\bibinfo  {journal} {JHEP}\ }\textbf {\bibinfo {volume} {08}},\ \bibinfo {pages} {135} (\bibinfo {year} {2017}{\natexlab{b}})},\ \Eprint {http://arxiv.org/abs/1612.06446} {arXiv:1612.06446 [hep-th]} \BibitemShut {NoStop}%
\bibitem [{\citenamefont {Dong}\ \emph {et~al.}(2023)\citenamefont {Dong}, \citenamefont {He},\ and\ \citenamefont {Lin}}]{Dong:2022bta}%
  \BibitemOpen
  \bibfield  {author} {\bibinfo {author} {\bibfnamefont {J.}~\bibnamefont {Dong}}, \bibinfo {author} {\bibfnamefont {S.}~\bibnamefont {He}}, \ and\ \bibinfo {author} {\bibfnamefont {G.}~\bibnamefont {Lin}},\ }\href {\doibase 10.1007/JHEP02(2023)076} {\bibfield  {journal} {\bibinfo  {journal} {JHEP}\ }\textbf {\bibinfo {volume} {02}},\ \bibinfo {pages} {076} (\bibinfo {year} {2023})},\ \Eprint {http://arxiv.org/abs/2208.00592} {arXiv:2208.00592 [hep-th]} \BibitemShut {NoStop}%
\bibitem [{\citenamefont {Dong}\ \emph {et~al.}(2022)\citenamefont {Dong}, \citenamefont {He},\ and\ \citenamefont {Hou}}]{Dong:2021qai}%
  \BibitemOpen
  \bibfield  {author} {\bibinfo {author} {\bibfnamefont {J.}~\bibnamefont {Dong}}, \bibinfo {author} {\bibfnamefont {S.}~\bibnamefont {He}}, \ and\ \bibinfo {author} {\bibfnamefont {L.}~\bibnamefont {Hou}},\ }\href {\doibase 10.1103/PhysRevD.105.105007} {\bibfield  {journal} {\bibinfo  {journal} {Phys. Rev. D}\ }\textbf {\bibinfo {volume} {105}},\ \bibinfo {pages} {105007} (\bibinfo {year} {2022})},\ \Eprint {http://arxiv.org/abs/2111.10525} {arXiv:2111.10525 [hep-th]} \BibitemShut {NoStop}%
\bibitem [{\citenamefont {Cao}\ \emph {et~al.}(2025)\citenamefont {Cao}, \citenamefont {Dong}, \citenamefont {He},\ and\ \citenamefont {Zhu}}]{Cao:2024olg}%
  \BibitemOpen
  \bibfield  {author} {\bibinfo {author} {\bibfnamefont {Q.}~\bibnamefont {Cao}}, \bibinfo {author} {\bibfnamefont {J.}~\bibnamefont {Dong}}, \bibinfo {author} {\bibfnamefont {S.}~\bibnamefont {He}}, \ and\ \bibinfo {author} {\bibfnamefont {F.}~\bibnamefont {Zhu}},\ }\href {\doibase 10.1103/PhysRevD.111.065015} {\bibfield  {journal} {\bibinfo  {journal} {Phys. Rev. D}\ }\textbf {\bibinfo {volume} {111}},\ \bibinfo {pages} {065015} (\bibinfo {year} {2025})},\ \Eprint {http://arxiv.org/abs/2412.19629} {arXiv:2412.19629 [hep-th]} \BibitemShut {NoStop}%
\bibitem [{\citenamefont {Cachazo}\ \emph {et~al.}(2014{\natexlab{d}})\citenamefont {Cachazo}, \citenamefont {Mason},\ and\ \citenamefont {Skinner}}]{Cachazo:2012pz}%
  \BibitemOpen
  \bibfield  {author} {\bibinfo {author} {\bibfnamefont {F.}~\bibnamefont {Cachazo}}, \bibinfo {author} {\bibfnamefont {L.}~\bibnamefont {Mason}}, \ and\ \bibinfo {author} {\bibfnamefont {D.}~\bibnamefont {Skinner}},\ }\href {\doibase 10.3842/SIGMA.2014.051} {\bibfield  {journal} {\bibinfo  {journal} {SIGMA}\ }\textbf {\bibinfo {volume} {10}},\ \bibinfo {pages} {051} (\bibinfo {year} {2014}{\natexlab{d}})},\ \Eprint {http://arxiv.org/abs/1207.4712} {arXiv:1207.4712 [hep-th]} \BibitemShut {NoStop}%
\bibitem [{\citenamefont {Baadsgaard}\ \emph {et~al.}(2015)\citenamefont {Baadsgaard}, \citenamefont {Bjerrum-Bohr}, \citenamefont {Bourjaily},\ and\ \citenamefont {Damgaard}}]{Baadsgaard:2015voa}%
  \BibitemOpen
  \bibfield  {author} {\bibinfo {author} {\bibfnamefont {C.}~\bibnamefont {Baadsgaard}}, \bibinfo {author} {\bibfnamefont {N.~E.~J.}\ \bibnamefont {Bjerrum-Bohr}}, \bibinfo {author} {\bibfnamefont {J.~L.}\ \bibnamefont {Bourjaily}}, \ and\ \bibinfo {author} {\bibfnamefont {P.~H.}\ \bibnamefont {Damgaard}},\ }\href {\doibase 10.1007/JHEP09(2015)129} {\bibfield  {journal} {\bibinfo  {journal} {JHEP}\ }\textbf {\bibinfo {volume} {09}},\ \bibinfo {pages} {129} (\bibinfo {year} {2015})},\ \Eprint {http://arxiv.org/abs/1506.06137} {arXiv:1506.06137 [hep-th]} \BibitemShut {NoStop}%
\bibitem [{\citenamefont {Cao}\ \emph {et~al.}(2024{\natexlab{a}})\citenamefont {Cao}, \citenamefont {Dong}, \citenamefont {He},\ and\ \citenamefont {Shi}}]{Cao:2024gln}%
  \BibitemOpen
  \bibfield  {author} {\bibinfo {author} {\bibfnamefont {Q.}~\bibnamefont {Cao}}, \bibinfo {author} {\bibfnamefont {J.}~\bibnamefont {Dong}}, \bibinfo {author} {\bibfnamefont {S.}~\bibnamefont {He}}, \ and\ \bibinfo {author} {\bibfnamefont {C.}~\bibnamefont {Shi}},\ }\href@noop {} {\  (\bibinfo {year} {2024}{\natexlab{a}})},\ \Eprint {http://arxiv.org/abs/2403.08855} {arXiv:2403.08855 [hep-th]} \BibitemShut {NoStop}%
\bibitem [{\citenamefont {Cao}\ \emph {et~al.}(2024{\natexlab{b}})\citenamefont {Cao}, \citenamefont {Dong}, \citenamefont {He}, \citenamefont {Shi},\ and\ \citenamefont {Zhu}}]{Cao:2024qpp}%
  \BibitemOpen
  \bibfield  {author} {\bibinfo {author} {\bibfnamefont {Q.}~\bibnamefont {Cao}}, \bibinfo {author} {\bibfnamefont {J.}~\bibnamefont {Dong}}, \bibinfo {author} {\bibfnamefont {S.}~\bibnamefont {He}}, \bibinfo {author} {\bibfnamefont {C.}~\bibnamefont {Shi}}, \ and\ \bibinfo {author} {\bibfnamefont {F.}~\bibnamefont {Zhu}},\ }\href {\doibase 10.1007/JHEP09(2024)049} {\bibfield  {journal} {\bibinfo  {journal} {JHEP}\ }\textbf {\bibinfo {volume} {09}},\ \bibinfo {pages} {049} (\bibinfo {year} {2024}{\natexlab{b}})},\ \Eprint {http://arxiv.org/abs/2406.03838} {arXiv:2406.03838 [hep-th]} \BibitemShut {NoStop}%
\bibitem [{\citenamefont {Arkani-Hamed}\ \emph {et~al.}(2023{\natexlab{a}})\citenamefont {Arkani-Hamed}, \citenamefont {Cao}, \citenamefont {Dong}, \citenamefont {Figueiredo},\ and\ \citenamefont {He}}]{Arkani-Hamed:2023swr}%
  \BibitemOpen
  \bibfield  {author} {\bibinfo {author} {\bibfnamefont {N.}~\bibnamefont {Arkani-Hamed}}, \bibinfo {author} {\bibfnamefont {Q.}~\bibnamefont {Cao}}, \bibinfo {author} {\bibfnamefont {J.}~\bibnamefont {Dong}}, \bibinfo {author} {\bibfnamefont {C.}~\bibnamefont {Figueiredo}}, \ and\ \bibinfo {author} {\bibfnamefont {S.}~\bibnamefont {He}},\ }\href@noop {} {\  (\bibinfo {year} {2023}{\natexlab{a}})},\ \Eprint {http://arxiv.org/abs/2312.16282} {arXiv:2312.16282 [hep-th]} \BibitemShut {NoStop}%
\bibitem [{\citenamefont {Rodina}(2025)}]{Rodina:2024yfc}%
  \BibitemOpen
  \bibfield  {author} {\bibinfo {author} {\bibfnamefont {L.}~\bibnamefont {Rodina}},\ }\href {\doibase 10.1103/PhysRevLett.134.031601} {\bibfield  {journal} {\bibinfo  {journal} {Phys. Rev. Lett.}\ }\textbf {\bibinfo {volume} {134}},\ \bibinfo {pages} {031601} (\bibinfo {year} {2025})},\ \Eprint {http://arxiv.org/abs/2406.04234} {arXiv:2406.04234 [hep-th]} \BibitemShut {NoStop}%
\bibitem [{\citenamefont {Zhou}(2025)}]{Zhou:2024ddy}%
  \BibitemOpen
  \bibfield  {author} {\bibinfo {author} {\bibfnamefont {K.}~\bibnamefont {Zhou}},\ }\href {\doibase 10.1007/JHEP03(2025)154} {\bibfield  {journal} {\bibinfo  {journal} {JHEP}\ }\textbf {\bibinfo {volume} {03}},\ \bibinfo {pages} {154} (\bibinfo {year} {2025})},\ \Eprint {http://arxiv.org/abs/2411.07944} {arXiv:2411.07944 [hep-th]} \BibitemShut {NoStop}%
\bibitem [{\citenamefont {Cachazo}\ \emph {et~al.}(2022)\citenamefont {Cachazo}, \citenamefont {Early},\ and\ \citenamefont {Gim\'enez~Umbert}}]{Cachazo:2021wsz}%
  \BibitemOpen
  \bibfield  {author} {\bibinfo {author} {\bibfnamefont {F.}~\bibnamefont {Cachazo}}, \bibinfo {author} {\bibfnamefont {N.}~\bibnamefont {Early}}, \ and\ \bibinfo {author} {\bibfnamefont {B.}~\bibnamefont {Gim\'enez~Umbert}},\ }\href {\doibase 10.1007/JHEP08(2022)252} {\bibfield  {journal} {\bibinfo  {journal} {JHEP}\ }\textbf {\bibinfo {volume} {08}},\ \bibinfo {pages} {252} (\bibinfo {year} {2022})},\ \Eprint {http://arxiv.org/abs/2112.14191} {arXiv:2112.14191 [hep-th]} \BibitemShut {NoStop}%
\bibitem [{\citenamefont {Arkani-Hamed}\ and\ \citenamefont {Figueiredo}(2024)}]{Arkani-Hamed:2024fyd}%
  \BibitemOpen
  \bibfield  {author} {\bibinfo {author} {\bibfnamefont {N.}~\bibnamefont {Arkani-Hamed}}\ and\ \bibinfo {author} {\bibfnamefont {C.}~\bibnamefont {Figueiredo}},\ }\href@noop {} {\  (\bibinfo {year} {2024})},\ \Eprint {http://arxiv.org/abs/2405.09608} {arXiv:2405.09608 [hep-th]} \BibitemShut {NoStop}%
\bibitem [{\citenamefont {Zhang}(2024)}]{Zhang:2024iun}%
  \BibitemOpen
  \bibfield  {author} {\bibinfo {author} {\bibfnamefont {Y.}~\bibnamefont {Zhang}},\ }\href@noop {} {\  (\bibinfo {year} {2024})},\ \Eprint {http://arxiv.org/abs/2406.08969} {arXiv:2406.08969 [hep-th]} \BibitemShut {NoStop}%
\bibitem [{\citenamefont {Zhang}(2025)}]{Zhang:2024efe}%
  \BibitemOpen
  \bibfield  {author} {\bibinfo {author} {\bibfnamefont {Y.}~\bibnamefont {Zhang}},\ }\href {\doibase 10.1007/JHEP02(2025)074} {\bibfield  {journal} {\bibinfo  {journal} {JHEP}\ }\textbf {\bibinfo {volume} {02}},\ \bibinfo {pages} {074} (\bibinfo {year} {2025})},\ \Eprint {http://arxiv.org/abs/2412.15198} {arXiv:2412.15198 [hep-th]} \BibitemShut {NoStop}%
\bibitem [{\citenamefont {Gim\'enez~Umbert}\ and\ \citenamefont {Sturmfels}(2025)}]{GimenezUmbert:2025ech}%
  \BibitemOpen
  \bibfield  {author} {\bibinfo {author} {\bibfnamefont {B.}~\bibnamefont {Gim\'enez~Umbert}}\ and\ \bibinfo {author} {\bibfnamefont {B.}~\bibnamefont {Sturmfels}},\ }\href@noop {} {\  (\bibinfo {year} {2025})},\ \Eprint {http://arxiv.org/abs/2501.10805} {arXiv:2501.10805 [hep-th]} \BibitemShut {NoStop}%
\bibitem [{Note4()}]{Note4}%
  \BibitemOpen
  \bibinfo {note} {One of the subset $A$ or $B$ can be empty.}\BibitemShut {Stop}%
\bibitem [{Note5()}]{Note5}%
  \BibitemOpen
  \bibinfo {note} {See the ref~\cite {Stieberger:2009hq,Stieberger:2015vya} for the detailed derivation and discussion.}\BibitemShut {Stop}%
\bibitem [{\citenamefont {Cachazo}\ \emph {et~al.}(2015)\citenamefont {Cachazo}, \citenamefont {He},\ and\ \citenamefont {Yuan}}]{Cachazo:2014xea}%
  \BibitemOpen
  \bibfield  {author} {\bibinfo {author} {\bibfnamefont {F.}~\bibnamefont {Cachazo}}, \bibinfo {author} {\bibfnamefont {S.}~\bibnamefont {He}}, \ and\ \bibinfo {author} {\bibfnamefont {E.~Y.}\ \bibnamefont {Yuan}},\ }\href {\doibase 10.1007/JHEP07(2015)149} {\bibfield  {journal} {\bibinfo  {journal} {JHEP}\ }\textbf {\bibinfo {volume} {07}},\ \bibinfo {pages} {149} (\bibinfo {year} {2015})},\ \Eprint {http://arxiv.org/abs/1412.3479} {arXiv:1412.3479 [hep-th]} \BibitemShut {NoStop}%
\bibitem [{\citenamefont {Cheung}\ \emph {et~al.}(2018)\citenamefont {Cheung}, \citenamefont {Shen},\ and\ \citenamefont {Wen}}]{Cheung:2017ems}%
  \BibitemOpen
  \bibfield  {author} {\bibinfo {author} {\bibfnamefont {C.}~\bibnamefont {Cheung}}, \bibinfo {author} {\bibfnamefont {C.-H.}\ \bibnamefont {Shen}}, \ and\ \bibinfo {author} {\bibfnamefont {C.}~\bibnamefont {Wen}},\ }\href {\doibase 10.1007/JHEP02(2018)095} {\bibfield  {journal} {\bibinfo  {journal} {JHEP}\ }\textbf {\bibinfo {volume} {02}},\ \bibinfo {pages} {095} (\bibinfo {year} {2018})},\ \Eprint {http://arxiv.org/abs/1705.03025} {arXiv:1705.03025 [hep-th]} \BibitemShut {NoStop}%
\bibitem [{\citenamefont {He}\ \emph {et~al.}(2021)\citenamefont {He}, \citenamefont {Hou}, \citenamefont {Tian},\ and\ \citenamefont {Zhang}}]{He:2021lro}%
  \BibitemOpen
  \bibfield  {author} {\bibinfo {author} {\bibfnamefont {S.}~\bibnamefont {He}}, \bibinfo {author} {\bibfnamefont {L.}~\bibnamefont {Hou}}, \bibinfo {author} {\bibfnamefont {J.}~\bibnamefont {Tian}}, \ and\ \bibinfo {author} {\bibfnamefont {Y.}~\bibnamefont {Zhang}},\ }\href {\doibase 10.1007/JHEP08(2021)118} {\bibfield  {journal} {\bibinfo  {journal} {JHEP}\ }\textbf {\bibinfo {volume} {08}},\ \bibinfo {pages} {118} (\bibinfo {year} {2021})},\ \bibinfo {note} {[Erratum: JHEP 06, 037 (2022)]},\ \Eprint {http://arxiv.org/abs/2103.15810} {arXiv:2103.15810 [hep-th]} \BibitemShut {NoStop}%
\bibitem [{\citenamefont {Ramond}(1971)}]{Ramond:1971gb}%
  \BibitemOpen
  \bibfield  {author} {\bibinfo {author} {\bibfnamefont {P.}~\bibnamefont {Ramond}},\ }\href {\doibase 10.1103/PhysRevD.3.2415} {\bibfield  {journal} {\bibinfo  {journal} {Phys. Rev. D}\ }\textbf {\bibinfo {volume} {3}},\ \bibinfo {pages} {2415} (\bibinfo {year} {1971})}\BibitemShut {NoStop}%
\bibitem [{\citenamefont {Neveu}\ and\ \citenamefont {Schwarz}(1971)}]{Neveu:1971rx}%
  \BibitemOpen
  \bibfield  {author} {\bibinfo {author} {\bibfnamefont {A.}~\bibnamefont {Neveu}}\ and\ \bibinfo {author} {\bibfnamefont {J.~H.}\ \bibnamefont {Schwarz}},\ }\href {\doibase 10.1016/0550-3213(71)90448-2} {\bibfield  {journal} {\bibinfo  {journal} {Nucl. Phys. B}\ }\textbf {\bibinfo {volume} {31}},\ \bibinfo {pages} {86} (\bibinfo {year} {1971})}\BibitemShut {NoStop}%
\bibitem [{\citenamefont {Berkovits}(2000)}]{Berkovits:2000fe}%
  \BibitemOpen
  \bibfield  {author} {\bibinfo {author} {\bibfnamefont {N.}~\bibnamefont {Berkovits}},\ }\href {\doibase 10.1088/1126-6708/2000/04/018} {\bibfield  {journal} {\bibinfo  {journal} {JHEP}\ }\textbf {\bibinfo {volume} {04}},\ \bibinfo {pages} {018} (\bibinfo {year} {2000})},\ \Eprint {http://arxiv.org/abs/hep-th/0001035} {arXiv:hep-th/0001035} \BibitemShut {NoStop}%
\bibitem [{\citenamefont {Berkovits}(2004)}]{Berkovits:2004px}%
  \BibitemOpen
  \bibfield  {author} {\bibinfo {author} {\bibfnamefont {N.}~\bibnamefont {Berkovits}},\ }\href {\doibase 10.1088/1126-6708/2004/09/047} {\bibfield  {journal} {\bibinfo  {journal} {JHEP}\ }\textbf {\bibinfo {volume} {09}},\ \bibinfo {pages} {047} (\bibinfo {year} {2004})},\ \Eprint {http://arxiv.org/abs/hep-th/0406055} {arXiv:hep-th/0406055} \BibitemShut {NoStop}%
\bibitem [{\citenamefont {Berkovits}(2005)}]{Berkovits:2005bt}%
  \BibitemOpen
  \bibfield  {author} {\bibinfo {author} {\bibfnamefont {N.}~\bibnamefont {Berkovits}},\ }\href {\doibase 10.1088/1126-6708/2005/10/089} {\bibfield  {journal} {\bibinfo  {journal} {JHEP}\ }\textbf {\bibinfo {volume} {10}},\ \bibinfo {pages} {089} (\bibinfo {year} {2005})},\ \Eprint {http://arxiv.org/abs/hep-th/0509120} {arXiv:hep-th/0509120} \BibitemShut {NoStop}%
\bibitem [{\citenamefont {Engelund}\ and\ \citenamefont {Roiban}(2013)}]{Engelund:2012re}%
  \BibitemOpen
  \bibfield  {author} {\bibinfo {author} {\bibfnamefont {O.~T.}\ \bibnamefont {Engelund}}\ and\ \bibinfo {author} {\bibfnamefont {R.}~\bibnamefont {Roiban}},\ }\href {\doibase 10.1007/JHEP03(2013)172} {\bibfield  {journal} {\bibinfo  {journal} {JHEP}\ }\textbf {\bibinfo {volume} {03}},\ \bibinfo {pages} {172} (\bibinfo {year} {2013})},\ \Eprint {http://arxiv.org/abs/1209.0227} {arXiv:1209.0227 [hep-th]} \BibitemShut {NoStop}%
\bibitem [{\citenamefont {Ahmed}\ and\ \citenamefont {Dhani}(2019)}]{Ahmed:2019upm}%
  \BibitemOpen
  \bibfield  {author} {\bibinfo {author} {\bibfnamefont {T.}~\bibnamefont {Ahmed}}\ and\ \bibinfo {author} {\bibfnamefont {P.~K.}\ \bibnamefont {Dhani}},\ }\href {\doibase 10.1007/JHEP05(2019)066} {\bibfield  {journal} {\bibinfo  {journal} {JHEP}\ }\textbf {\bibinfo {volume} {05}},\ \bibinfo {pages} {066} (\bibinfo {year} {2019})},\ \Eprint {http://arxiv.org/abs/1901.08086} {arXiv:1901.08086 [hep-th]} \BibitemShut {NoStop}%
\bibitem [{\citenamefont {Ahmed}\ \emph {et~al.}(2020)\citenamefont {Ahmed}, \citenamefont {Banerjee}, \citenamefont {Chakraborty}, \citenamefont {Dhani},\ and\ \citenamefont {Ravindran}}]{Ahmed:2019yjt}%
  \BibitemOpen
  \bibfield  {author} {\bibinfo {author} {\bibfnamefont {T.}~\bibnamefont {Ahmed}}, \bibinfo {author} {\bibfnamefont {P.}~\bibnamefont {Banerjee}}, \bibinfo {author} {\bibfnamefont {A.}~\bibnamefont {Chakraborty}}, \bibinfo {author} {\bibfnamefont {P.~K.}\ \bibnamefont {Dhani}}, \ and\ \bibinfo {author} {\bibfnamefont {V.}~\bibnamefont {Ravindran}},\ }\href {\doibase 10.1103/PhysRevD.102.061701} {\bibfield  {journal} {\bibinfo  {journal} {Phys. Rev. D}\ }\textbf {\bibinfo {volume} {102}},\ \bibinfo {pages} {061701} (\bibinfo {year} {2020})},\ \Eprint {http://arxiv.org/abs/1911.11886} {arXiv:1911.11886 [hep-th]} \BibitemShut {NoStop}%
\bibitem [{\citenamefont {Stieberger}(2021)}]{Stieberger:2021daa}%
  \BibitemOpen
  \bibfield  {author} {\bibinfo {author} {\bibfnamefont {S.}~\bibnamefont {Stieberger}},\ }\href@noop {} {\  (\bibinfo {year} {2021})},\ \Eprint {http://arxiv.org/abs/2105.06888} {arXiv:2105.06888 [hep-th]} \BibitemShut {NoStop}%
\bibitem [{\citenamefont {Arkani-Hamed}\ \emph {et~al.}(2023{\natexlab{b}})\citenamefont {Arkani-Hamed}, \citenamefont {Frost}, \citenamefont {Salvatori}, \citenamefont {Plamondon},\ and\ \citenamefont {Thomas}}]{Arkani-Hamed:2023lbd}%
  \BibitemOpen
  \bibfield  {author} {\bibinfo {author} {\bibfnamefont {N.}~\bibnamefont {Arkani-Hamed}}, \bibinfo {author} {\bibfnamefont {H.}~\bibnamefont {Frost}}, \bibinfo {author} {\bibfnamefont {G.}~\bibnamefont {Salvatori}}, \bibinfo {author} {\bibfnamefont {P.-G.}\ \bibnamefont {Plamondon}}, \ and\ \bibinfo {author} {\bibfnamefont {H.}~\bibnamefont {Thomas}},\ }\href@noop {} {\  (\bibinfo {year} {2023}{\natexlab{b}})},\ \Eprint {http://arxiv.org/abs/2309.15913} {arXiv:2309.15913 [hep-th]} \BibitemShut {NoStop}%
\bibitem [{\citenamefont {Arkani-Hamed}\ \emph {et~al.}(2023{\natexlab{c}})\citenamefont {Arkani-Hamed}, \citenamefont {Frost}, \citenamefont {Salvatori}, \citenamefont {Plamondon},\ and\ \citenamefont {Thomas}}]{Arkani-Hamed:2023mvg}%
  \BibitemOpen
  \bibfield  {author} {\bibinfo {author} {\bibfnamefont {N.}~\bibnamefont {Arkani-Hamed}}, \bibinfo {author} {\bibfnamefont {H.}~\bibnamefont {Frost}}, \bibinfo {author} {\bibfnamefont {G.}~\bibnamefont {Salvatori}}, \bibinfo {author} {\bibfnamefont {P.-G.}\ \bibnamefont {Plamondon}}, \ and\ \bibinfo {author} {\bibfnamefont {H.}~\bibnamefont {Thomas}},\ }\href@noop {} {\  (\bibinfo {year} {2023}{\natexlab{c}})},\ \Eprint {http://arxiv.org/abs/2311.09284} {arXiv:2311.09284 [hep-th]} \BibitemShut {NoStop}%
\bibitem [{\citenamefont {Arkani-Hamed}\ \emph {et~al.}(2023{\natexlab{d}})\citenamefont {Arkani-Hamed}, \citenamefont {Cao}, \citenamefont {Dong}, \citenamefont {Figueiredo},\ and\ \citenamefont {He}}]{Arkani-Hamed:2023jry}%
  \BibitemOpen
  \bibfield  {author} {\bibinfo {author} {\bibfnamefont {N.}~\bibnamefont {Arkani-Hamed}}, \bibinfo {author} {\bibfnamefont {Q.}~\bibnamefont {Cao}}, \bibinfo {author} {\bibfnamefont {J.}~\bibnamefont {Dong}}, \bibinfo {author} {\bibfnamefont {C.}~\bibnamefont {Figueiredo}}, \ and\ \bibinfo {author} {\bibfnamefont {S.}~\bibnamefont {He}},\ }\href@noop {} {\  (\bibinfo {year} {2023}{\natexlab{d}})},\ \Eprint {http://arxiv.org/abs/2401.00041} {arXiv:2401.00041 [hep-th]} \BibitemShut {NoStop}%
\bibitem [{\citenamefont {Arkani-Hamed}\ \emph {et~al.}(2024{\natexlab{a}})\citenamefont {Arkani-Hamed}, \citenamefont {Cao}, \citenamefont {Dong}, \citenamefont {Figueiredo},\ and\ \citenamefont {He}}]{Arkani-Hamed:2024nhp}%
  \BibitemOpen
  \bibfield  {author} {\bibinfo {author} {\bibfnamefont {N.}~\bibnamefont {Arkani-Hamed}}, \bibinfo {author} {\bibfnamefont {Q.}~\bibnamefont {Cao}}, \bibinfo {author} {\bibfnamefont {J.}~\bibnamefont {Dong}}, \bibinfo {author} {\bibfnamefont {C.}~\bibnamefont {Figueiredo}}, \ and\ \bibinfo {author} {\bibfnamefont {S.}~\bibnamefont {He}},\ }\href@noop {} {\  (\bibinfo {year} {2024}{\natexlab{a}})},\ \Eprint {http://arxiv.org/abs/2401.05483} {arXiv:2401.05483 [hep-th]} \BibitemShut {NoStop}%
\bibitem [{\citenamefont {Arkani-Hamed}\ \emph {et~al.}(2024{\natexlab{b}})\citenamefont {Arkani-Hamed}, \citenamefont {Figueiredo}, \citenamefont {Frost},\ and\ \citenamefont {Salvatori}}]{Arkani-Hamed:2024vna}%
  \BibitemOpen
  \bibfield  {author} {\bibinfo {author} {\bibfnamefont {N.}~\bibnamefont {Arkani-Hamed}}, \bibinfo {author} {\bibfnamefont {C.}~\bibnamefont {Figueiredo}}, \bibinfo {author} {\bibfnamefont {H.}~\bibnamefont {Frost}}, \ and\ \bibinfo {author} {\bibfnamefont {G.}~\bibnamefont {Salvatori}},\ }\href@noop {} {\  (\bibinfo {year} {2024}{\natexlab{b}})},\ \Eprint {http://arxiv.org/abs/2402.06719} {arXiv:2402.06719 [hep-th]} \BibitemShut {NoStop}%
\bibitem [{\citenamefont {Arkani-Hamed}\ \emph {et~al.}(2024{\natexlab{c}})\citenamefont {Arkani-Hamed}, \citenamefont {Cao}, \citenamefont {Dong}, \citenamefont {Figueiredo},\ and\ \citenamefont {He}}]{Arkani-Hamed:2024tzl}%
  \BibitemOpen
  \bibfield  {author} {\bibinfo {author} {\bibfnamefont {N.}~\bibnamefont {Arkani-Hamed}}, \bibinfo {author} {\bibfnamefont {Q.}~\bibnamefont {Cao}}, \bibinfo {author} {\bibfnamefont {J.}~\bibnamefont {Dong}}, \bibinfo {author} {\bibfnamefont {C.}~\bibnamefont {Figueiredo}}, \ and\ \bibinfo {author} {\bibfnamefont {S.}~\bibnamefont {He}},\ }\href@noop {} {\  (\bibinfo {year} {2024}{\natexlab{c}})},\ \Eprint {http://arxiv.org/abs/2408.11891} {arXiv:2408.11891 [hep-th]} \BibitemShut {NoStop}%
\bibitem [{\citenamefont {Arkani-Hamed}\ \emph {et~al.}(2024{\natexlab{d}})\citenamefont {Arkani-Hamed}, \citenamefont {Figueiredo},\ and\ \citenamefont {Remmen}}]{Arkani-Hamed:2024nzc}%
  \BibitemOpen
  \bibfield  {author} {\bibinfo {author} {\bibfnamefont {N.}~\bibnamefont {Arkani-Hamed}}, \bibinfo {author} {\bibfnamefont {C.}~\bibnamefont {Figueiredo}}, \ and\ \bibinfo {author} {\bibfnamefont {G.~N.}\ \bibnamefont {Remmen}},\ }\href@noop {} {\  (\bibinfo {year} {2024}{\natexlab{d}})},\ \Eprint {http://arxiv.org/abs/2412.20639} {arXiv:2412.20639 [hep-th]} \BibitemShut {NoStop}%
\bibitem [{\citenamefont {Arkani-Hamed}\ \emph {et~al.}(2024{\natexlab{e}})\citenamefont {Arkani-Hamed}, \citenamefont {Frost},\ and\ \citenamefont {Salvatori}}]{Arkani-Hamed:2024pzc}%
  \BibitemOpen
  \bibfield  {author} {\bibinfo {author} {\bibfnamefont {N.}~\bibnamefont {Arkani-Hamed}}, \bibinfo {author} {\bibfnamefont {H.}~\bibnamefont {Frost}}, \ and\ \bibinfo {author} {\bibfnamefont {G.}~\bibnamefont {Salvatori}},\ }\href@noop {} {\  (\bibinfo {year} {2024}{\natexlab{e}})},\ \Eprint {http://arxiv.org/abs/2412.21027} {arXiv:2412.21027 [hep-th]} \BibitemShut {NoStop}%
\bibitem [{\citenamefont {De}\ \emph {et~al.}(2024)\citenamefont {De}, \citenamefont {Pokraka}, \citenamefont {Skowronek}, \citenamefont {Spradlin},\ and\ \citenamefont {Volovich}}]{De:2024wsy}%
  \BibitemOpen
  \bibfield  {author} {\bibinfo {author} {\bibfnamefont {S.}~\bibnamefont {De}}, \bibinfo {author} {\bibfnamefont {A.}~\bibnamefont {Pokraka}}, \bibinfo {author} {\bibfnamefont {M.}~\bibnamefont {Skowronek}}, \bibinfo {author} {\bibfnamefont {M.}~\bibnamefont {Spradlin}}, \ and\ \bibinfo {author} {\bibfnamefont {A.}~\bibnamefont {Volovich}},\ }\href {\doibase 10.1007/JHEP09(2024)160} {\bibfield  {journal} {\bibinfo  {journal} {JHEP}\ }\textbf {\bibinfo {volume} {09}},\ \bibinfo {pages} {160} (\bibinfo {year} {2024})},\ \Eprint {http://arxiv.org/abs/2406.04411} {arXiv:2406.04411 [hep-th]} \BibitemShut {NoStop}%
\bibitem [{\citenamefont {Cao}\ and\ \citenamefont {Zhu}(2025)}]{Cao:2025mlt}%
  \BibitemOpen
  \bibfield  {author} {\bibinfo {author} {\bibfnamefont {Q.}~\bibnamefont {Cao}}\ and\ \bibinfo {author} {\bibfnamefont {F.}~\bibnamefont {Zhu}},\ }\href@noop {} {\  (\bibinfo {year} {2025})},\ \Eprint {http://arxiv.org/abs/2503.15860} {arXiv:2503.15860 [hep-th]} \BibitemShut {NoStop}%
\bibitem [{\citenamefont {Arkani-Hamed}\ \emph {et~al.}(2023{\natexlab{e}})\citenamefont {Arkani-Hamed}, \citenamefont {He}, \citenamefont {Lam},\ and\ \citenamefont {Thomas}}]{Arkani-Hamed:2019plo}%
  \BibitemOpen
  \bibfield  {author} {\bibinfo {author} {\bibfnamefont {N.}~\bibnamefont {Arkani-Hamed}}, \bibinfo {author} {\bibfnamefont {S.}~\bibnamefont {He}}, \bibinfo {author} {\bibfnamefont {T.}~\bibnamefont {Lam}}, \ and\ \bibinfo {author} {\bibfnamefont {H.}~\bibnamefont {Thomas}},\ }\href {\doibase 10.1103/PhysRevD.107.066015} {\bibfield  {journal} {\bibinfo  {journal} {Phys. Rev. D}\ }\textbf {\bibinfo {volume} {107}},\ \bibinfo {pages} {066015} (\bibinfo {year} {2023}{\natexlab{e}})},\ \Eprint {http://arxiv.org/abs/1912.11764} {arXiv:1912.11764 [hep-th]} \BibitemShut {NoStop}%
\bibitem [{\citenamefont {Arkani-Hamed}\ \emph {et~al.}(2022)\citenamefont {Arkani-Hamed}, \citenamefont {He}, \citenamefont {Salvatori},\ and\ \citenamefont {Thomas}}]{Arkani-Hamed:2019vag}%
  \BibitemOpen
  \bibfield  {author} {\bibinfo {author} {\bibfnamefont {N.}~\bibnamefont {Arkani-Hamed}}, \bibinfo {author} {\bibfnamefont {S.}~\bibnamefont {He}}, \bibinfo {author} {\bibfnamefont {G.}~\bibnamefont {Salvatori}}, \ and\ \bibinfo {author} {\bibfnamefont {H.}~\bibnamefont {Thomas}},\ }\href {\doibase 10.1007/JHEP11(2022)049} {\bibfield  {journal} {\bibinfo  {journal} {JHEP}\ }\textbf {\bibinfo {volume} {11}},\ \bibinfo {pages} {049} (\bibinfo {year} {2022})},\ \Eprint {http://arxiv.org/abs/1912.12948} {arXiv:1912.12948 [hep-th]} \BibitemShut {NoStop}%
\bibitem [{\citenamefont {Lin}\ and\ \citenamefont {Yang}(2022)}]{Lin:2021pne}%
  \BibitemOpen
  \bibfield  {author} {\bibinfo {author} {\bibfnamefont {G.}~\bibnamefont {Lin}}\ and\ \bibinfo {author} {\bibfnamefont {G.}~\bibnamefont {Yang}},\ }\href {\doibase 10.1103/PhysRevLett.129.251601} {\bibfield  {journal} {\bibinfo  {journal} {Phys. Rev. Lett.}\ }\textbf {\bibinfo {volume} {129}},\ \bibinfo {pages} {251601} (\bibinfo {year} {2022})},\ \Eprint {http://arxiv.org/abs/2111.12719} {arXiv:2111.12719 [hep-th]} \BibitemShut {NoStop}%
\bibitem [{\citenamefont {Guillen}\ \emph {et~al.}(2021)\citenamefont {Guillen}, \citenamefont {Johansson}, \citenamefont {Jusinskas},\ and\ \citenamefont {Schlotterer}}]{Guillen:2021mwp}%
  \BibitemOpen
  \bibfield  {author} {\bibinfo {author} {\bibfnamefont {M.}~\bibnamefont {Guillen}}, \bibinfo {author} {\bibfnamefont {H.}~\bibnamefont {Johansson}}, \bibinfo {author} {\bibfnamefont {R.~L.}\ \bibnamefont {Jusinskas}}, \ and\ \bibinfo {author} {\bibfnamefont {O.}~\bibnamefont {Schlotterer}},\ }\href {\doibase 10.1103/PhysRevLett.127.051601} {\bibfield  {journal} {\bibinfo  {journal} {Phys. Rev. Lett.}\ }\textbf {\bibinfo {volume} {127}},\ \bibinfo {pages} {051601} (\bibinfo {year} {2021})},\ \Eprint {http://arxiv.org/abs/2104.03314} {arXiv:2104.03314 [hep-th]} \BibitemShut {NoStop}%
\bibitem [{\citenamefont {Kashyap}\ \emph {et~al.}(2023)\citenamefont {Kashyap}, \citenamefont {Mafra}, \citenamefont {Verma},\ and\ \citenamefont {Ypanaque}}]{Kashyap:2023cdi}%
  \BibitemOpen
  \bibfield  {author} {\bibinfo {author} {\bibfnamefont {S.~P.}\ \bibnamefont {Kashyap}}, \bibinfo {author} {\bibfnamefont {C.~R.}\ \bibnamefont {Mafra}}, \bibinfo {author} {\bibfnamefont {M.}~\bibnamefont {Verma}}, \ and\ \bibinfo {author} {\bibfnamefont {L.~A.}\ \bibnamefont {Ypanaque}},\ }\href@noop {} {\  (\bibinfo {year} {2023})},\ \Eprint {http://arxiv.org/abs/2311.12100} {arXiv:2311.12100 [hep-th]} \BibitemShut {NoStop}%
\bibitem [{\citenamefont {Kashyap}\ \emph {et~al.}(2025)\citenamefont {Kashyap}, \citenamefont {Mafra}, \citenamefont {Verma},\ and\ \citenamefont {Ypanaqu\'e}}]{Kashyap:2024qor}%
  \BibitemOpen
  \bibfield  {author} {\bibinfo {author} {\bibfnamefont {S.~P.}\ \bibnamefont {Kashyap}}, \bibinfo {author} {\bibfnamefont {C.~R.}\ \bibnamefont {Mafra}}, \bibinfo {author} {\bibfnamefont {M.}~\bibnamefont {Verma}}, \ and\ \bibinfo {author} {\bibfnamefont {L.}~\bibnamefont {Ypanaqu\'e}},\ }\href {\doibase 10.1007/JHEP02(2025)215} {\bibfield  {journal} {\bibinfo  {journal} {JHEP}\ }\textbf {\bibinfo {volume} {02}},\ \bibinfo {pages} {215} (\bibinfo {year} {2025})},\ \Eprint {http://arxiv.org/abs/2407.02436} {arXiv:2407.02436 [hep-th]} \BibitemShut {NoStop}%
\bibitem [{\citenamefont {Mafra}(2024)}]{Mafra:2024fiy}%
  \BibitemOpen
  \bibfield  {author} {\bibinfo {author} {\bibfnamefont {C.~R.}\ \bibnamefont {Mafra}},\ }\href {\doibase 10.1007/JHEP11(2024)045} {\bibfield  {journal} {\bibinfo  {journal} {JHEP}\ }\textbf {\bibinfo {volume} {11}},\ \bibinfo {pages} {045} (\bibinfo {year} {2024})},\ \Eprint {http://arxiv.org/abs/2407.11849} {arXiv:2407.11849 [hep-th]} \BibitemShut {NoStop}%
\end{thebibliography}%



\end{document}